\documentclass[12pt]{article}
\input{amssym.def}
\input{amssym}
\usepackage[dvips]{color}
\usepackage{epsfig}
\usepackage{subcaption}
\usepackage{hyperref}
\normalbaselines
\makeatletter
\@addtoreset{equation}{section}
\makeatother

\def\be{\begin{equation}}
\def\ee{\end{equation}}
\def\ba{\begin{eqnarray}}
\def\ea{\end{eqnarray}}

\newcommand{\rf}[1]{(\ref{#1})}
\def\bra#1{\langle #1|}
\def\ket#1{|#1\rangle}

\def\scp{\scriptsize}

\begin{document}

\title
{
Critical phases in the raise and peel model }
\author{D. A. C. Jara$^{1,2}$, and   F. C. Alcaraz$^1$
\\[5mm] {\small\it$^{1}$
Instituto de F\'{\i}sica de S\~{a}o Carlos, Universidade de S\~{a}o Paulo, Caixa Postal 369,} \\
{\small\it 13560-590, S\~{a}o Carlos, SP, Brazil} \\
 \small\it$^{2}$Faculdad de Ingenieria, Universidad de La Sabana, 
A.A. 53753, Chia,Colombia}
\date{\today}
\maketitle
\footnotetext[1]{\tt dacj1984@gmail.com}
\footnotetext[2]{\tt alcaraz@if.sc.usp.br}


\begin{abstract}
The raise and peel model (RPM) is a nonlocal stochastic model describing the 
space and time fluctuations of an evolving one dimensional interface. Its 
relevant parameter $u$ is the ratio  between the rates of local adsorption
and nonlocal desorption processes (avalanches) processes. The model at $u=1$ give us the 
first example of a conformally invariant stochastic model. For small values 
$u<u_0$ the model is known to be noncritical, while for $u>u_0$ it is critical. 
Although previous studies indicate that $u_0=1$ the determination of $u_0$ 
with a reasonable precision is still missing. By calculating the structure 
 function of the height profiles in the reciprocal space we confirm with 
good precision that indeed $u_0=1$. 
We establish that at the 
conformal invariant point $u=1$ the RPM has a roughness transition with 
dynamical and roughness critical exponents $z=1$ and $\alpha=0$, respectively. 
For $u>1$ the model is critical with an $u$-dependent dynamical critical 
exponent $z(u)$ that tends towards zero as $u\to \infty$. However at $1/u=0$ 
the RPM is exactly mapped into the totally asymmetric exclusion problem 
(TASEP). This last model is known to be noncritical (critical) for open (periodic) boundary conditions. Our studies indicate that the RPM as $u \to \infty$, 
due to its nonlocal dynamics processes, has the same large-distance physics 
no matter what boundary condition we chose. For $u>1$, our analysis show that 
differently from previous predictions, the region is composed by two distinct 
critical phases. For $u\leq u < u_c\approx 40$ the height profiles are rough 
($\alpha = \alpha(u) >0$), and for $u>u_c$ the height profiles are flat at 
large distances ($\alpha = \alpha(u) <0$). We also observed that in both 
critical phases ($u>1$) the RPM at short length scales, has an effective 
behavior in the Kardar-Parisi-Zhang (KPZ) critical universality class, that 
is not the true behavior of the system at large length scales.

\end{abstract}

\section{ Introduction} \label{sect1}

The raise and peel model (RPM) on its original formulation 
\cite{RPM1,RPM2,RPM3} is a stochastic model describing the time and space 
fluctuations of restricted solid-on-solid (RSOS) one dimensional profiles. 
They define an interface separating a solid phase from a rarefied gas of 
tiles, and change as the tiles coming from the rarefied gas phase reach the 
interface. The tiles can be locally absorbed  (raise) or can trigger a nonlocal 
desorption of tiles (peel) in the surface of the solid phase. The model is defined in terms 
of a free parameter $u$ given by the ratio between the adsorption and 
desorption rates. The RSOS fluctuating profiles can also represent the 
configurations of excluded volume particles in a discrete lattice. The RPM in 
this case \cite{RPM4} gives a generalization of the asymmetric exclusion 
problem (ASEP) \cite{ASEP}, where the excluded volume particles are allowed 
to have local jumps to the sites on the left and nonlocal ones to the sites 
on the right. The parameter $u$ defines the anisotropy left/right of the 
possible motions. 

The model at $u=1$ (equal rates of adsorption and desorption) is special. Its 
time-evolution operator (Hamiltonian) is exactly integrable and give us the 
first example of a stochastic model conformally invariant.  Its 
Hamiltonian is given in terms of the generators of the Temperley-Lieb algebra. 
The model, for the case of open boundaries, can be mapped onto the spin-zero 
sector of the XXZ quantum chain with the quantum $U_q(Sl(2))$ symmetry with 
$q=e^{i\frac{\pi}{3}}$ \cite{RPM1,RPM2,RPM3}. In the case of periodic 
boundaries it  is related to a XXZ quantum chain with twisted boundary 
condition \cite{RPM4} (twisted angle $\phi=\frac{2\pi}{3}$). 

For $u\neq 1$ all the known results of the RPM comes from numerical analysis. 
In the original presentation of the model \cite{RPM1} it was not clear if 
a phase transition exists for $u=u_0\approx 0.5$ separating a massive 
phase ($u<u_0$) from a critical phase ($u>u_0$). The value $u_0\approx 0.5$ 
was suggested from the mass-gap amplitude crossings of the eigenenergies of 
the associated Hamiltonian, with  open (free) boundary 
conditions.  In \cite{RPM2}  numerical studies, based on Monte Carlo 
simulations, also with the RPM with open boundaries, indicate that $u_0=1$. 
Moreover for $u>1$ the model is in a self-organized criticality (SOC) phase 
where the dynamical critical exponent decreases from the value $z=1$, at the 
conformal invariant point $u=1$, to the value $z=0$ when $u\to \infty$ (no 
desorption).

In this paper we present an extensive numerical study of the phase diagram of 
the RPM. The original motivation of  these calculations is due 
to the exact connection \cite{Jara} of the RPM with no desorption ($1/u=0$) 
with the totally asymmetric exclusion problem (TASEP) \cite{TASEP1,TASEP2}. 
The TASEP is critical ($z=3/2$) or not depending if the boundary condition 
is periodic or not. Since all the previous calculations of the dynamical 
critical exponents were done only in the case of open boundaries is it 
important to verify if indeed the critical behavior (critical exponents) are 
the same for both boundaries, differently of the limiting case $1/u=0$. 
Measuring several distinct observables we were able to confirm that indeed 
the massive phase ends up at the critical point $u=u_0=1$. By calculating the
roughness of  the profiles we then verified that the 
RPM has a roughness transition at $u=1$, with roughness critical exponent 
$\alpha=0$.

for $u>1$ we obtain some unexpected results that were not observed in the 
previous calculations of the RPM with open ends \cite{RPM2,RPM3}. Differently 
from previous studies, where it was expected a single critical phase, our 
results indicate the existence of two distinct critical phases: 
$1\leq u \leq u_c$ and $u>u_c$, with $u_c \approx 40$. The intermediate phase 
($1\leq u \leq u_c$) is rough with a roughness critical exponent 
$\alpha = \alpha(u) \geq 0$. The second phase ($u>u_c$) is critical and 
flat at large length scales. In this last phase the average height, as the lattice 
size increases, reaches a limit in the case of open boundaries. In the 
periodic case the average height of the surface increases with time, reaching 
a stationary finite velocity.  This velocity 
corresponds to a limiting current  in the particle formulation of the RPM \cite{RPM4}.

We also verified that the time-evolution in both critical phases exhibits the 
phenomena of {\it critical initial slip } \cite{slip}. A phenomena that 
induces the appearance, for a quite large time interval, of 
an effective exponent 
that depends on the particular configuration where the system starts  its 
evolution. This phenomena produces difficulties in the evaluation of the 
dynamical critical exponent by using the Family-Visek scaling, as reported 
in earlier calculations of the RPM \cite{RPM2}. To avoid this effect we should 
calculate observables directly in the stationary state. A quite reasonable 
assumption of the scaling behavior of the height profiles at the stationary 
regime indicates a possible way to evaluate the dynamical critical exponent. 

Our results indicate that independently of the boundary condition, as 
$u\to \infty$, the dynamical critical exponent goes to zero. This is distinct 
from the case where we set $1/u=0$ (no desorption). In this limiting case we 
recover the TASEP, a stochastic model that is critical and belongs to the 
Kardar-Parisi-Zhang (KPZ) universality class \cite{KPZ}, only if the 
boundary conditions are closed (periodic). In order to see how the height 
profiles change as we approach the limiting case $u=u_0=1$ and $u\to \infty$ 
we calculate the spatial structure function of the height profiles. The 
results show us clearly crossover effects to the KPZ behavior as $u$ 
increases, and allow us to understand the distinct behavior in both 
critical phases. 

The paper is organized as follows. In the next section, in order to set the 
notations we present the RPM as well the observables that will be considered 
along the paper. In section 3 we give the results that confirm that for 
$u<u_0$ the model is noncritical, being $u_0=1$ (conformally invariant point) 
 the critical point, 
where the model is conformally invariant. The section 4 is devoted to the 
evaluation of the critical exponents for $u>u_0=1$, by using several distinct 
methods. In section 5 we calculate the spatial structure function of the 
height profiles. The results indicate the existence of two distinct 
critical phases. Finally in section 6 we present our concluding remarks.

\section{ Description of the raise and peel model} \label{sect2}

  The stochastic model RPM has a free parameter $u$ and is already described 
in several papers \cite{RPM1,RPM2,RPM3}. The model, for the special value 
$u=1$, gives  the first example of a conformally invariant stochastic model 
(central charge $c=0$). Its time-evolution operator 
(Hamiltonian) is related to the exact integrable XXZ quantum chain with 
anisotropy $\Delta=-1/2$ and special boundaries \cite{RPM1,RPM2,RPM3}. At this 
conformal invariant point the model is also interesting, from the mathematical 
point of view since, as observed by Razumov and Stroganov \cite{Razumov}, the 
probability distribution of the system's configurations is related to the 
interesting problem of enumerating alternated sign matrices. 

\begin{figure}
\centering
\includegraphics[angle=0,width=0.50\textwidth] {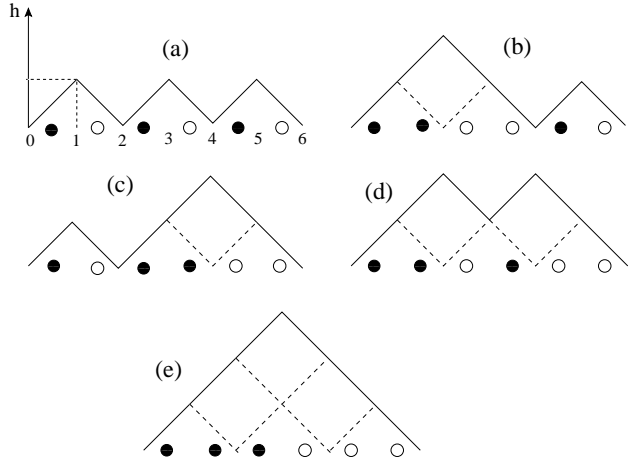}
\caption{
The five  configurations of the RPM for a lattice with $L=6$ sites 
and open boundary conditions.
The configurations in the particle-vacancy 
representation are also shown: particles and vacancies are denoted by 
full and empty circles.}
\label{fig1}
\end{figure}

\begin{figure}
\centering
\includegraphics[angle=0,width=0.45\textwidth] {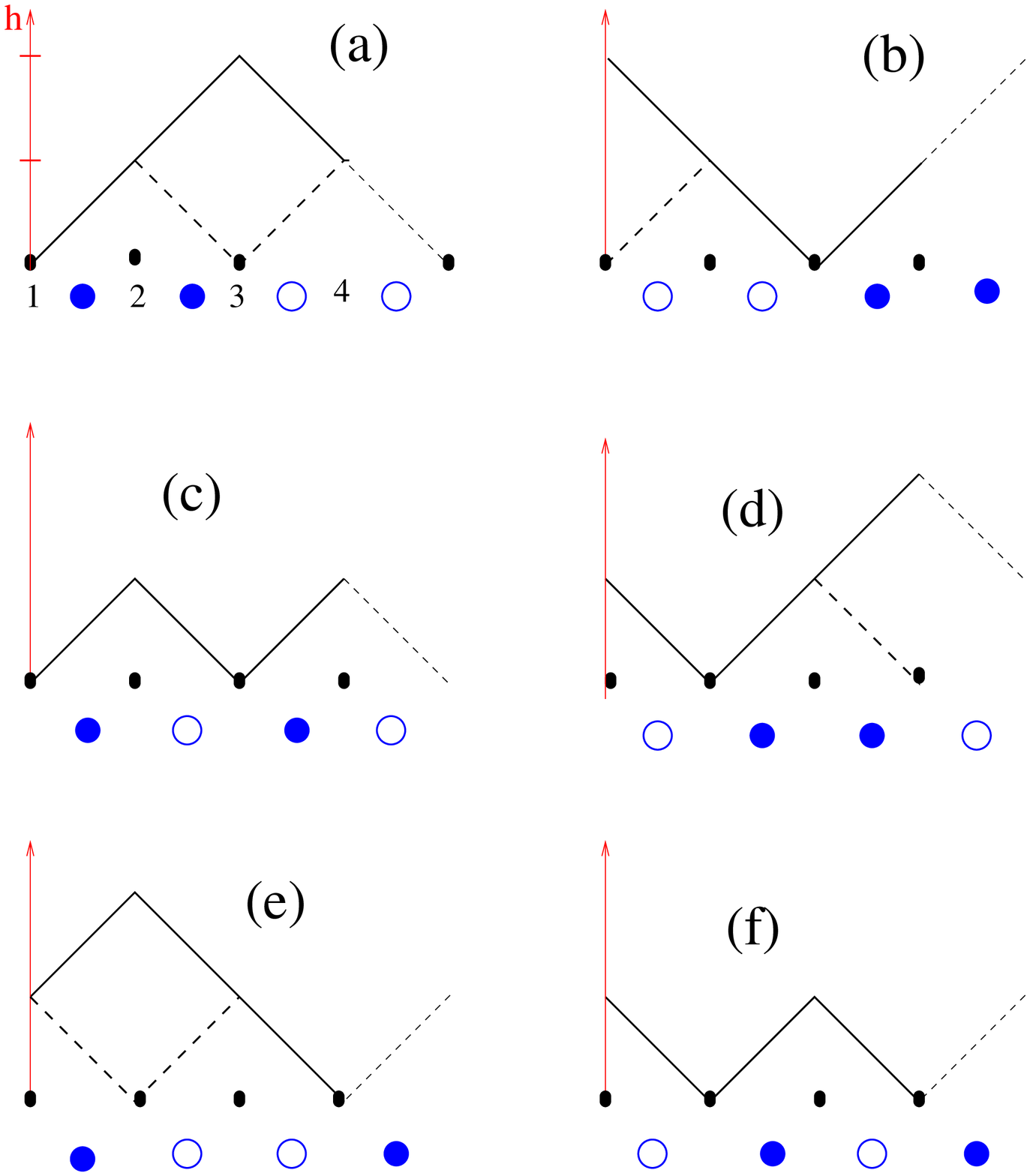}
\caption{
The 6 possible profiles for the RPM with periodic boundaries and 
$L=4$ sites.
The configurations in the particle-vacancy 
representation are also shown: particles and vacancies are denoted by 
full and empty circles.}
\label{fig2}
\end{figure}

We are going to describe the model in two distinct, but equivalent, stochastic 
bases (configuration's space). The first one, that we named the height 
representation basis, is given in terms of Dyck paths. These paths are 
defined in terms of integer heights $\{h_i\}$ obeying the restricted 
solid-on-solid (SOS) rules:
\be \label{e2.1}
h_{i+1}-h_i = \pm 1.
\ee
For open systems $i=0,1,2,\ldots,L$ and $h_0=h_L=0$, and for the periodic 
ones $i=1,2,\ldots,L$ and $h_i=h_{i+L}$. We consider in this paper $L$ as an 
even number. There are \\
$L!/\{(L+2)([L/2]!)^2\}$ configurations for the open
systems and $L!/([L/2]!)^2$ for the periodic case. The second basis, we 
formulated the RPM, is the one we call particle-vacancy basis \cite{RPM4}. 
It is given by the configurations of $L/2$ excluded volume  particles 
and $L/2$ vacancies. The configurations in this last basis are obtained from 
the height representation one by inserting particles or vacancies at the 
links $(i,i+1)$, depending if $h_{i+1} > h_i$ or $h_{i+1}<h_i$, respectively. 
In Figs.~\ref{fig1} and \ref{fig2} we show the configurations in both basis 
for the $L=6$ model with open boundaries and for the $L=4$ model with 
periodic boundaries, respectively.


In the height basis we can visualize the stochastic evolution of the RPM 
by considering the profiles $\{h_i\}$ as an one dimensional surface 
separating  a solid  from a rarefied gas phase of tilted tiles (see Fig.~\ref{fig3}). The profiles changes due to the hits of tiles coming from the 
rarefied gas phase, according to the following rules: 

\begin{figure}
\centering
\includegraphics[angle=0,width=0.7\textwidth] {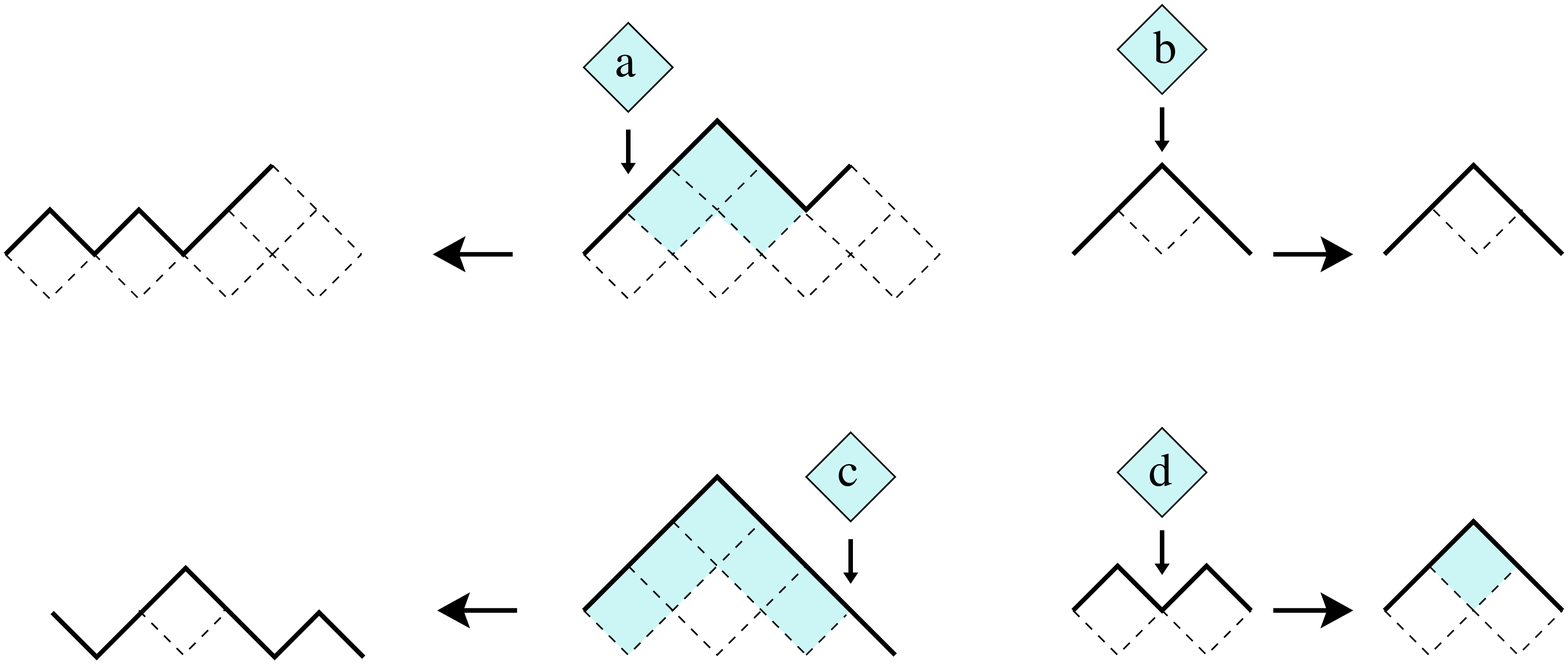}
\caption{
The dynamical processes in the RPM. The tilted tile in the gaseous phase
hit a positive slope (a), a peak in (b), a negative slope in (c)
and a valley in (d) (see the text).}
\label{fig3}
\end{figure}

During a short time interval $\Delta t$ at most one tile from the gaseous 
phase reaches  the surface. With a probability $p=\Delta t/L$ a tile reaches
 the site $i$ of the surface (see Fig.~\ref{fig3}). The allowed motions depend 
on the local heights ($h_{i-1},h_i,h_{i+1}$).

{\it i)} If $h_{i-1}<h_i>h_{i+1}$ the tile reaches a local peak (case (b) of 
Fig.~\ref{fig3}). The tile is reflected with no changes in the surface profile.

{\it ii)} If $h_{i-1}>h_i<h_{i+1}$ the tile reaches a local minimum 
(case (d) of 
Fig.~\ref{fig3}). With a probability $p_a$, proportional to the absorption 
rate $W_a$, the tile is added to the surface ($h_i \to h_i+2$) and with a 
probability $1-p_a$ the tile is reflected, with no change in the surface 
profile.

{\it iii)} If $h_{i-1}<h_i<h_{i+1}$ the tile reaches a positive slope 
(case (a) of Fig.~\ref{fig3}). With a probability $p_d$ proportional to the 
desorption rate $W_d$, the tile is reflected after desorbing ($h_i \to h_i-2$) 
a layer of ($b-1$) tiles from the segment $\{h_i\}$ ($j=i+1,\ldots,i+b-1$), 
where $h_j>h_i=h_{i+b}$. With the probability $1-p_d$ the tile is reflected 
and no changes happen in the surface.

{\it iv)} If $h_{i-1}>h_i>h_{i+1}$ the tile reaches a negative slope 
(case (c) of Fig.~\ref{fig3}). With a probability $p_d$, proportional to the 
desorption rate $W_d$, the tile is reflected after desorbing ($h_j \to h_j-2$) 
a layer of ($b-1$) tiles from the segment $\{h_j\}$ ($j=i-b+1,\ldots,i-1$), where 
$h_j>h_i=h_{i-b}$. With a probability $1-p_d$ the tile is reflected and the 
surface is unchanged.

 The relevant parameter in the dynamics of the RPM is the ratio among the 
adsorbing and desorbing rates $u=W_a/W_d$. In a continuous time evolution 
the time fluctuations of the probabilities $P_c(t)$ of finding the system 
in a configuration $c$ is given by the master equation
\be \label{e2.2}
\frac{dP_c(t)}{dt} = \sum_{c'} H_{c,c'} P_{c'}(t),
\ee
where $H_{c,c'}$ are the matrix elements of the Hamiltonian governing the 
stochastic evolution. As an example, the Hamiltonian $H$ for the open chain 
with $L=6$ sites, connecting the five configurations of Fig.~\ref{fig1}, is 
given by
\ba \label{e2.3}
 H =-
\left( \begin{array}{c|rrrrr}
 & \bra{a} & \bra{b} & \bra{c} & \bra{d} &\bra{e} \\ \hline
\ket{a} &  -2u  & 2 & 2 &0 &2  \\
\ket{b} &  u  & -(u+2) & 0 &1 &0  \\
\ket{c} &  u  & 0 & -(u+2) &1 &0  \\
\ket{d} &  0  & u & u &-(u+2) &2  \\
\ket{e} & 0  & 0 & 0 &u &-4  
\end{array} \right) ,&&\nonumber \\
\ea 
where  we take $u_d=1$ and $u=u_a$. 

\begin{figure}
\centering
\includegraphics[angle=0,width=0.35\textwidth] {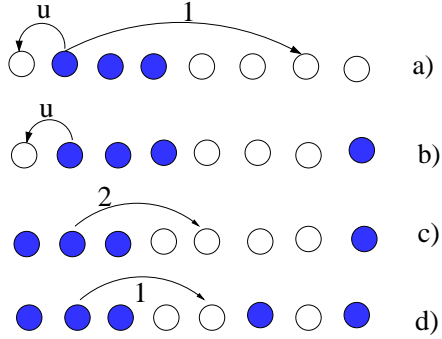}
\caption{
       a,b) Hopping rules if the particle is preceded by a
      vacancy. c,d) Hopping rules if the particle is preceded by
      another particle.}
\label{fig4}
\end{figure}

The dynamic rules for the RPM, in the particle-vacancy basis, follows from the 
correspondence of the configurations in this basis with those in the height 
basis. In this representation the particles (only the particles) can make 
jumps to the leftmost position with probability proportional to $W_a$, 
provide it is empty (has a vacancy), or may do a nonlocal jump to empty 
positions on its right, leaving a segment with equal number of particles and 
vacancies. The probability of these jumps to the right are proportional to the
desorption rate and depends also on the configuration of  the particles. 
 In Fig.~\ref{fig4} we give schematically the allowed motions of particles 
in the particle-vacancy basis (see also \cite{RPM4}). 

Due to the dynamics of the RPM in the case of open boundaries two 
configurations have special importance: the {\it substrate} and the 
{\it pyramid}
configurations. The substrate configuration is the one where $h_i=h_{i+2}$ 
($i=0,1,\ldots,L$) and no desorptions are allowed. In Fig.~\ref{fig1} it is the 
configuration (a). The pyramid configuration is the one where $h_i=i$ ($i=0,\ldots,L/2$) and $h_i = L-i$ ($i=\frac{L}{2}+1, \ldots,L$), and only desorptions 
may take place. In Fig.~\ref{fig1} is the configuration (e).

In the following sections, in order to characterize the phase diagram of the RPM, we are going to evaluate some observables. An important one is the 
average height of the profile at time $t$:
\be \label{e2.4}
h(L,(t) = \frac{1}{L} \sum_{i=1}^L < h_i(t)>,
\ee
where $<h_i(t)>$ is the average height at site $i=1,\ldots,L$ in the height 
representation. The Fourier transform of the heights 
\be \label{e2.5}
h(k,t)=\frac{1}{L^{1/2}} \sum_{j=1}^L (h_j(t) - h(L,t))\exp(ikj), 
\ee 
give us the structure function
\be \label{e2.6}
S(k,t) = <h(k,t)h(-k,t)>, \quad k=\frac{2\pi j}{L}\quad \quad (j=-L/2,\ldots,L/2),
\ee
that reveals the structure of the height profiles at the spatial length $\lambda=2\pi/k$.

In the open boundary case it is interesting to defined the contact points and 
the clusters. Contact points are the sites of a given profile where the height is 
zero and the profile makes a contact with the substrate at $h=0$, as for example the 
points 0, 4 and 6 of the configurations (b) of Fig.~\ref{fig1}. 
 The heights between two  consecutive contact points, with tiles added in the substrate is  defined as a cluster. In Fig.~\ref{fig1} 
there are one cluster in configurations ((b), (c), (d) and (e), and no cluster in the substrate configuration (a).

\section{The phase transition at $u=u_0=1$} \label{section3}

For $u<<1$ the RPM is clearly in a massive phase. The stationary asymptotic 
state (ground state of the Hamiltonian) is given basically by a combination 
of the substrate configuration and the ones obtained from the addition of 
 few tiles in the substrate. We have a combination of a large number of 
clusters with finite size, independent of the lattice size, for sufficiently large lattice sizes. As $u$ increases the characteristic size of the clusters 
increases and at $u=u_0$ it diverges with the lattice size. The precise 
determination of the critical point $u=u_0$, where the massive phase ends, is 
not simple. 

\begin{figure}
\centering
\includegraphics[angle=0,width=0.4\textwidth] {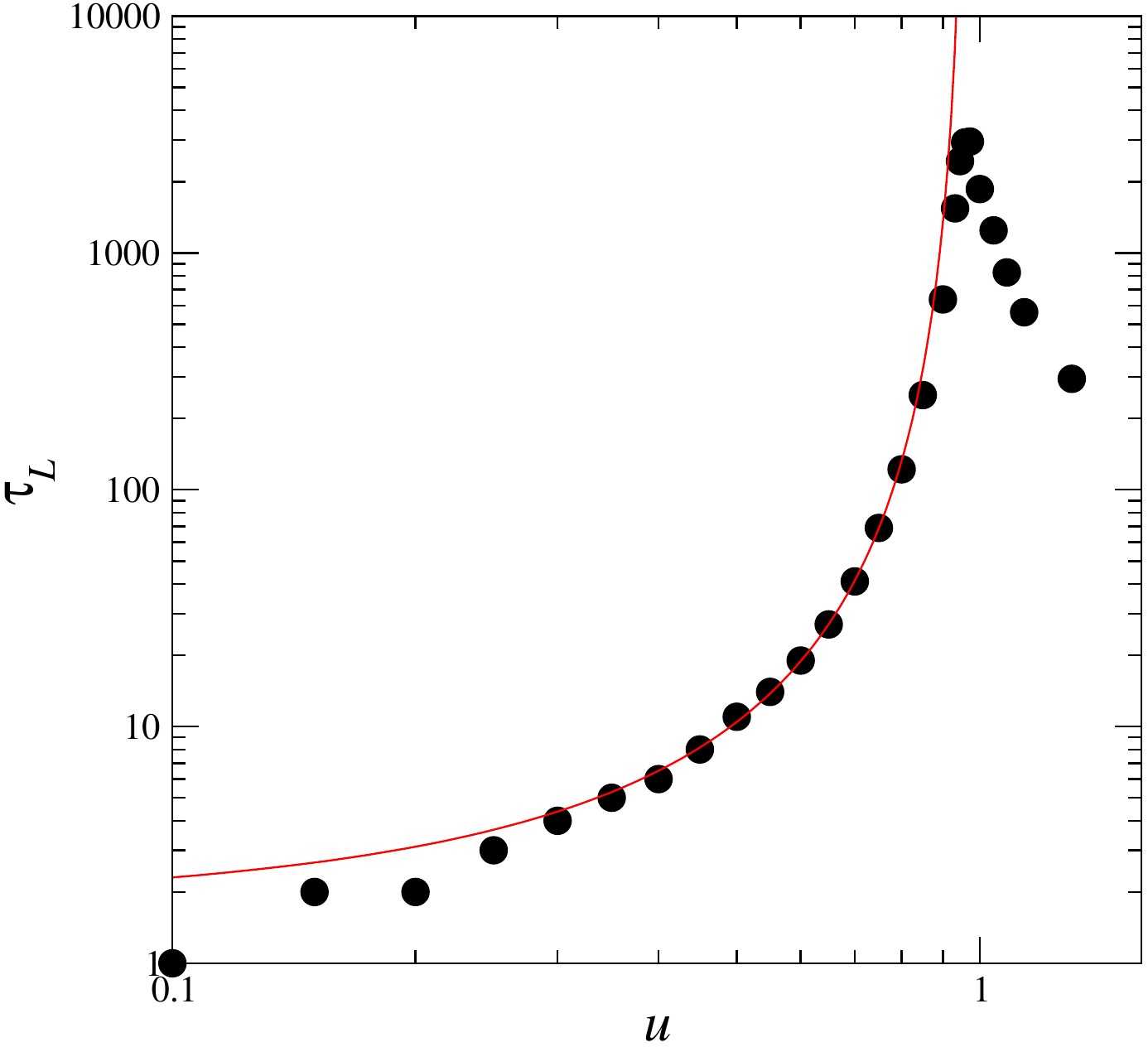}
\caption{
The characteristic time $\tau_L$ (units of Monte Carlos steps) necessary 
to reach the stationary state, as a function of $u$. The RPM has $L=16384$ 
sites, open ends, and the initial configuration is the substrate one.}
\label{fig5}
\end{figure}

In \cite{RPM1}, where the RPM was introduced, numerical calculations of the 
mass gap crossings of the Hamiltonian with small lattice sizes and open 
boundary conditions indicates that $u_0\approx 0.5$.  However the finite-size 
effects around $u\approx u_0$ are quite large. To illustrate them we show in 
Fig.~\ref{fig5} the characteristic time $\tau_L$, as a function of $u$, 
necessary for the system to reach the stationary state, by starting ($t=0$) 
with the substrate configuration of a lattice with $L=2^{14}=16384$ sites.
The fitting curve, shown in red in Fig.~\ref{fig5} indicates 
$\tau_L\sim (u_0-u)^{-\nu_t}$, with $u_0=0.96$ and $\nu_t=2.43$. Although 
the lattice size is relatively large, by increasing the lattice size for 
$L>2^{17}=131072$ we verified that the predicted value of $u_0$ increases. 
In fact a previous calculation \cite{RPM2}, based on the density of clusters, 
although considering a relatively  small lattice size ($L=2^{16}=65636$) indicates that 
$u_0=1$. 

Although we have done  measurements of several observables for large lattice 
sizes that confirms the phase transition at $u=u_0=1$, we are going to present
 in this section the measurements of only two of these observables, with the 
lattice taken with free boundaries. The first one is the average size of a
 cluster $CS$, that we define as:
\be \label{e3.1}
CS= \sum_{k=1}^{N_c} s_k \left(\frac{s_k}{N_{oc}}\right),
\ee
where $s_k$ is the number of tiles in the first row ($h=2$) of the $k$th 
cluster ($k=1,2,\ldots,N_c$), $N_c$ is the number of clusters and $N_{oc}= \sum_{k=1}^{N_c}s_k$ is the total number of tiles in the first row of the profile 
configuration. The quantity $CS$ can be interpreted as follows. If we chose 
a random site among the ones having tiles in the first row ($h_i\geq 2$), $CS$ 
gives the average size of the cluster where the chosen site belongs. 

\begin{figure}
\begin{subfigure}{.5\textwidth}
\centering
\includegraphics[angle=0,width=0.8\textwidth] {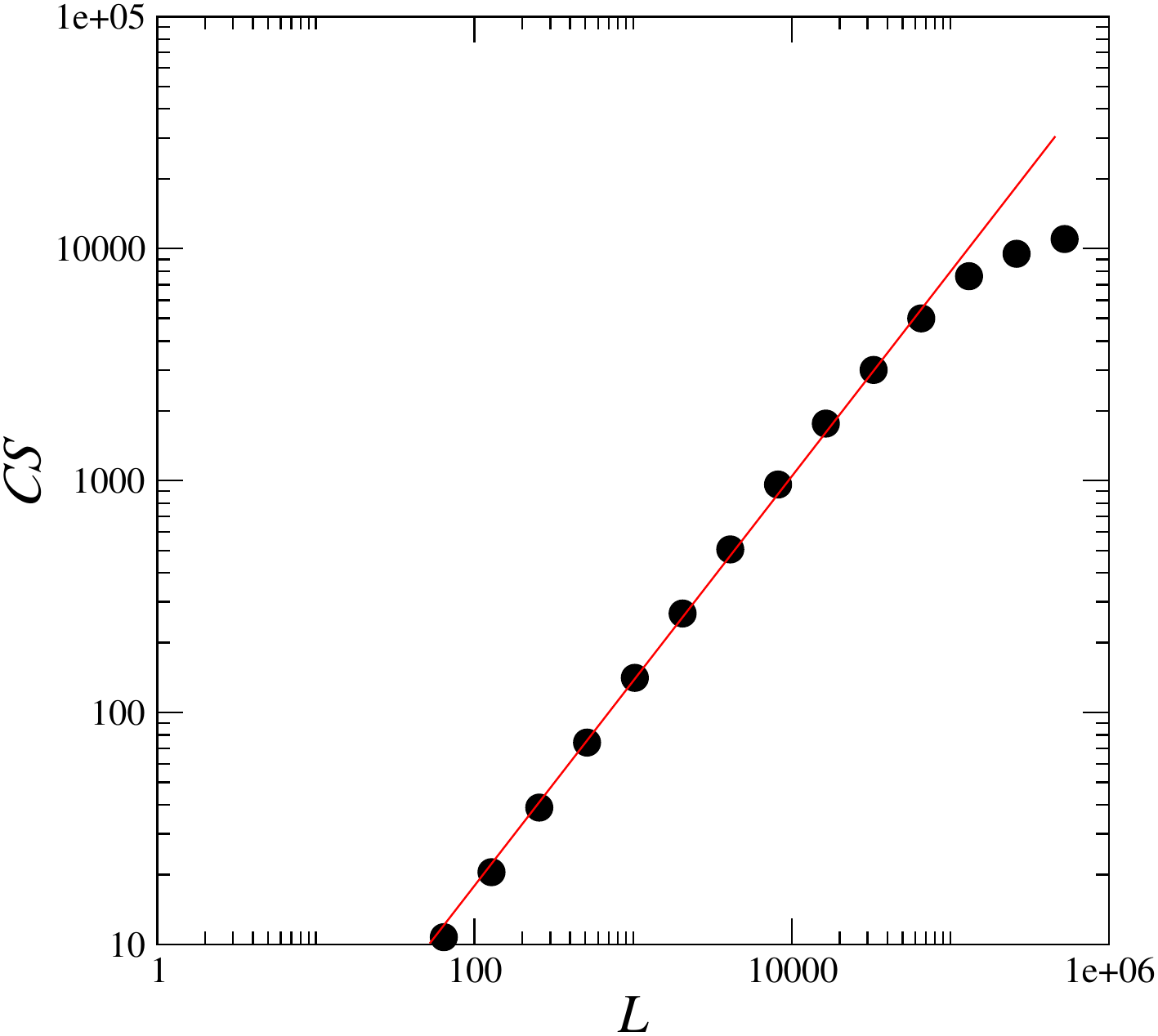}
\label{fig6a}
\end{subfigure}
\begin{subfigure}{.5\textwidth}
\centering
\includegraphics[angle=0,width=0.8\textwidth] {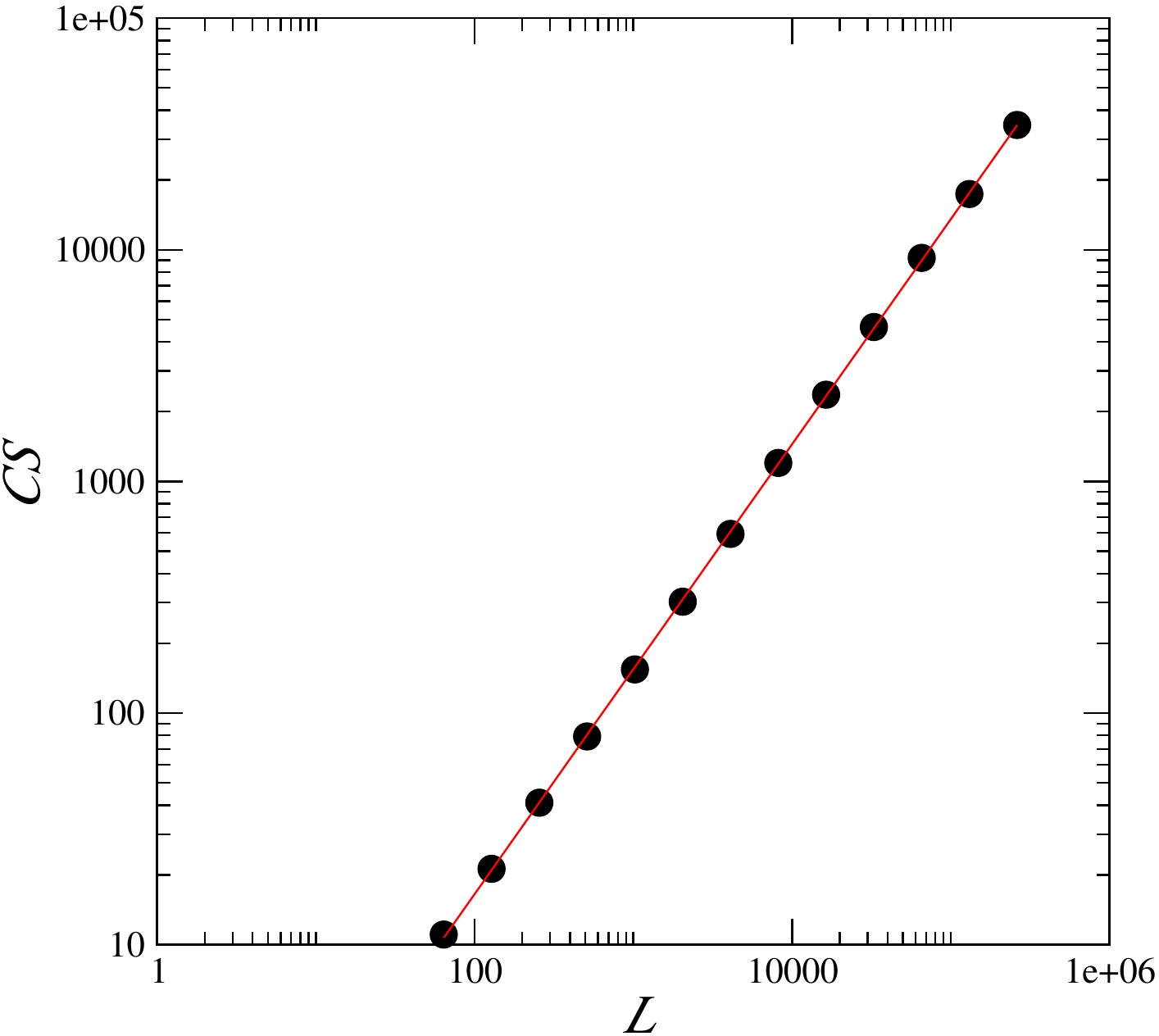}
\label{fig6b}
\end{subfigure}
\caption{
Average cluster size $CS$ (black dots), in the stationary state of the RPM
with open boundaries, as a function of the lattice size L. a) $u=0.98$, b) u=1. 
The red lines are guides to show the deviations from a straight line.}
\label{fig6}
\end{figure}

In Figs.~\ref{fig6}a and \ref{fig6}b we show $CS$ for lattice sizes up to 
$L=2^{19}=524,288$ for $u=0.98$ and $u=1$, respectively. We clearly see in 
Fig.~\ref{fig6}a that for $u=0.98$ the  lattice sizes $L>100000$ 
indicate a saturation value for a finite cluster size as $L\to \infty$, as 
we should expect in a massive phase. On the other hand at $u=1$, as shown 
in Fig.~\ref{fig6}b, the average cluster size diverges with the lattice size. 

\begin{figure}
\centering
\includegraphics[angle=0,width=0.4\textwidth] {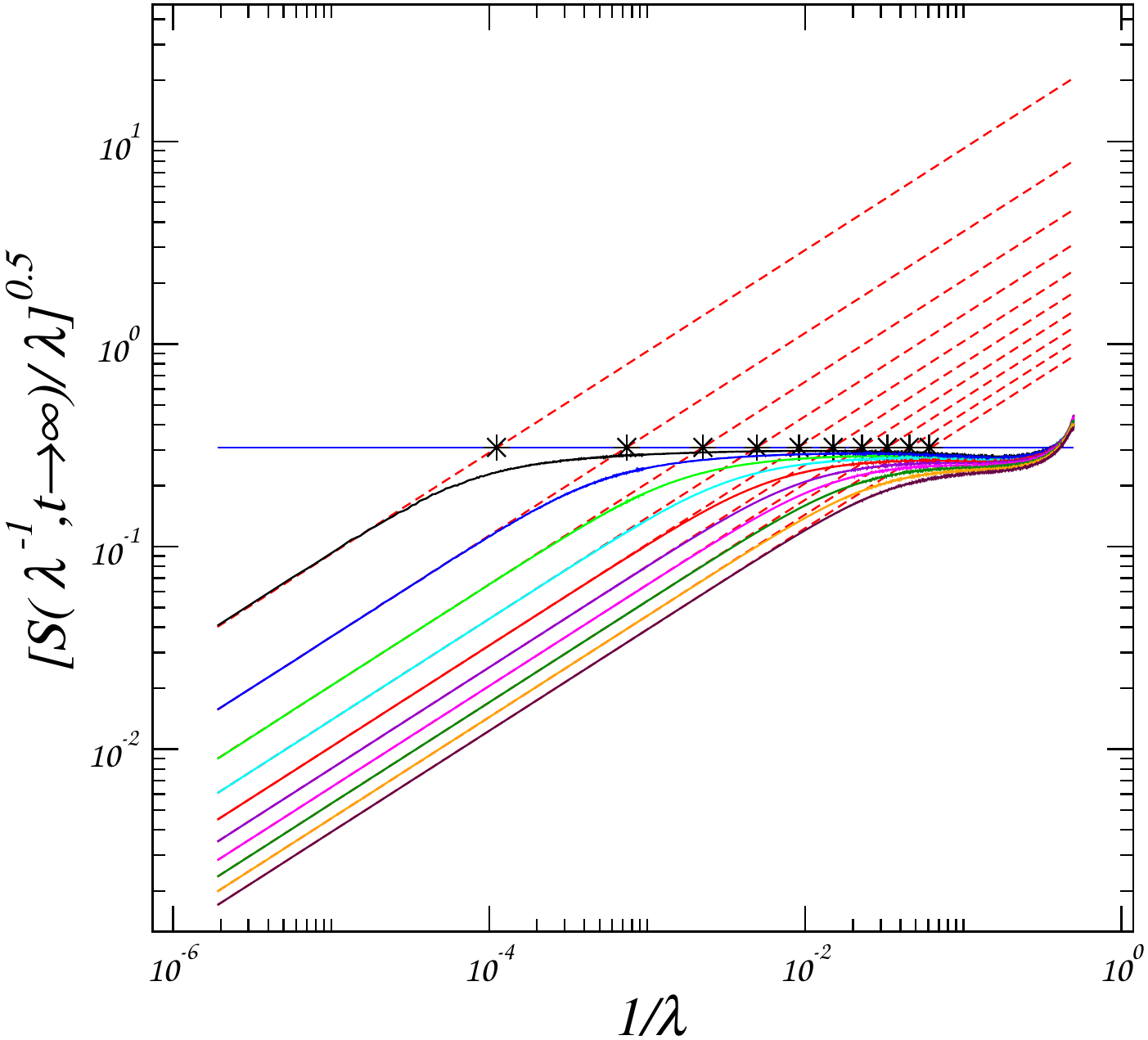}
\caption{
The structure function $S(1/\lambda,t\to \infty)$ of the RPM at the stationary 
regime for several values of $u$ and lattice size $L=524,382$. The values of 
$u$ from the top to the bottom are $0.95, 0.90,,\ldots, 0.55,0.50$.
. The red 
straight lines are obtained by fitting the curves, by considering small values 
of $\lambda$. The straight blue horizontal line is the  auxiliary line 
used to determine the crossing points ($*$) separating the initial and 
asymptotic regimes. } 
\label{fig7}
\end{figure}

The second observable we measured to confirm $u_0=1$ is the spatial structure 
function of the profiles at the stationary state ($t \to \infty$), defined in \rf{e2.5}-\rf{e2.6}. 
  For several values of $u$ the structure function 
$S(1/\lambda,t\to \infty)$, for the lattice size $L=524,382$ is shown in 
Fig.~\ref{fig7}. The values of $u$ in the figure are, from right to left, 
$u=0.50,0.55,\ldots,0.90,0.95$. We see in this figure that at short length 
scales, i. e., $\lambda \lesssim \lambda_1(u)$ the structure function behaves 
as $S(\lambda,t\to \infty) \sim \lambda$ (or 
$\sqrt{S(\lambda, t \to \infty)/\lambda} \sim $ const.), while for large 
scales $\lambda \gtrsim \lambda_1(u)$ it behaves as 
$S(1/\lambda, t \to \infty) \sim $ const.,  implying that the 
profiles are composed by finite-size clusters whose typical sizes increases 
with $u$. The constant behavior for $\sqrt{S(1/\lambda, t\to \infty)/\lambda}$ 
in Fig.~\ref{fig7}, for $\lambda <\lambda_1(u)$, i. e., 
$S(1/\lambda,t \to \infty) \sim \lambda$, is a consequence of a crossover 
effect due to the critical behavior at the conformal invariant point $u=1$, 
since as we will see in section 5 $S(1/\lambda,t \to \infty) \sim \lambda$ 
(see Eq.~\rf{e5.5}).

   Fig.~\ref{fig7} also explains the strong finite-size effects for 
 $u \gtrsim 0.95 $, as we saw in Fig.~\ref{fig6}a. For the model with $u=0.95$ we 
see an "effective critical" crossover effect up to $\lambda \sim 10^5$, 
implying that the massive behavior of the model can only be seen for lattice 
sizes $L\gtrsim 10^5$, in agreement with the results shown in Fig.~\ref{fig6}a.

\begin{figure}
\centering
\includegraphics[angle=0,width=0.4\textwidth] {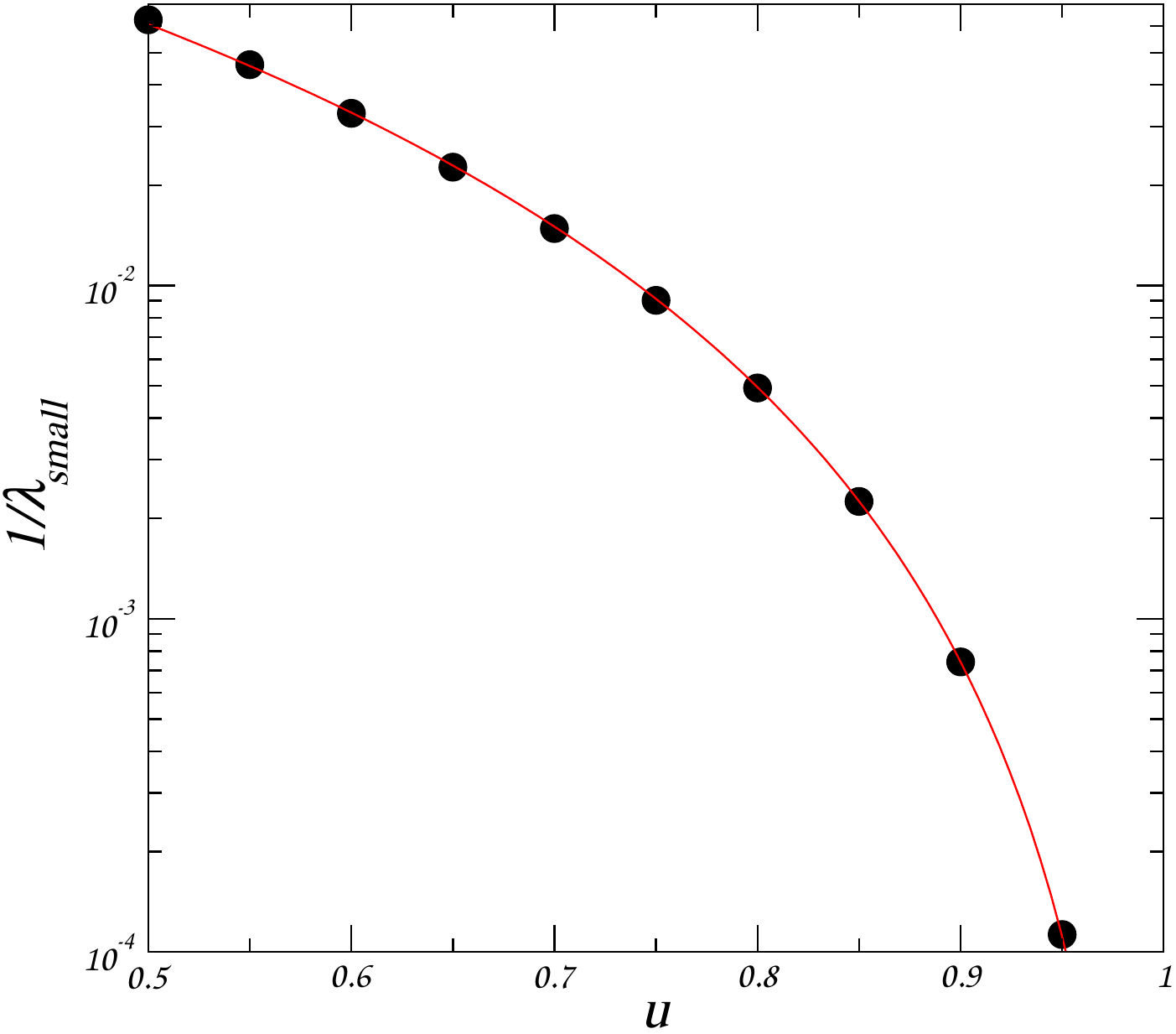}
\caption{ 
The crossing points $\lambda_{\scriptsize{\mbox{small}}}(u)$, represented 
as ($*$) in Fig.~\ref{fig7}, as a function of $u$. }
\label{fig8}
\end{figure}

In order to get a precise estimate of the critical point $u_0$ from 
Fig.~\ref{fig7} we consider the wavelength $\lambda_{\mbox{\small{small}}}(u)$ 
obtained  from the crossing of the two distinct asymptotic behaviors, 
namely, the one for  
$\lambda \to \infty$ and the one expected at $u=1$ 
(horizontal line in Fig.~i\ref{fig7}). Those are the points marked in 
(*) in Fig.~\ref{fig7}. In Fig.~\ref{fig8} we plot 
$\lambda_{\mbox{\small{small}}}(u)$ and we got the fit 
\be \label{e3.2}
1/\lambda_{\mbox{\small{small}}}(u) = a(u-u_0)^b, 
\ee
where $a=0.406(4)$, $u_0=1.000(1)$ and the exponent $b=2.738(7)$, 
confirming that $u_0=1$. 

\section{The critical exponents and phases for $u>1$} \label{sect4}

Previous numerical results for the RPM with open \cite{RPM1,RPM2} and 
periodic boundary conditions \cite{RPM4} indicate that for $u>1$ the model is 
in a critical regime. We are going to present in this section new numerical 
results that indicate that this critical regime is separated into two 
critical phases where the profiles exhibit distinct behavior at large scales. 

Previous evaluations \cite{RPM2} of the dynamical critical exponent $z$ for 
$u\ge 1$ indicate a continuous decrease as $u$ increases, tending to $z=0$ 
as $u \to \infty$. On the other hand, more recently \cite{Jara} it was 
observed that the RPM at the limiting case $1/u=0$ recovers exactly the 
totally asymmetric exclusion problem (TASEP). We can see this correspondence 
easily from the allowed processes in the particle-vacancy representation of 
the model. We can see in Fig.~\ref{fig4} that at this limit the particles 
can only move to the left, provide the leftmost site is empty. All the local 
and nonlocal jumps to the right are not allowed. The TASEP although critical 
for periodic chains is not critical for open boundaries. In the periodic 
critical case it belongs to the Kardar-Parisi-Zhang (KPZ) universality 
class \cite{KPZ} where the dynamical critical exponent  $z=3/2$ and the 
roughness critical exponent is $\alpha=1/2$. A natural question arises for the RPM: is this 
distinct behavior for different boundaries a characteristic of the $1/u=0$ 
limit, or we may have it also in the RPM for $u$ large, but finite? As we 
shall see our numerical results indicate that these boundary-dependent 
behaviors only happen at the singular point $1/u=0$. Once $1/u\neq 0$ the
allowed nonlocal jumps, that happen for both boundaries, make the model 
critical, with a small value for the critical exponent $z(u)\approx 0$, 
for $1/u \approx 0$, indicating the discontinuity 
$z(u\to \infty) - z(1/u=0) \neq 0$, for the case of periodic boundaries. 

A possible way to estimate the dynamical critical exponent $z=z(u)$ of the 
RPM is obtained from the leading finite-size behavior of the mass gaps 
associated to the excited eigenvalues  $E_n$ ($n=1,2,\ldots$) of its 
$L$-site Hamiltonian (see \rf{e2.3} for the case $L=6$ and open boundaries):
\be \label{e4.1}
E_n= \frac{A_n}{L^z} + o(L^{-z}), \quad n=1,2,\ldots .
\ee

Applying the power method to estimate the lowest eigenstates of the model 
with free boundaries we were able to calculate the gaps $E_1$ and $E_2$ up 
to $L=30$. In table 1 we show the estimated values $z(E_1)$ 
and $z(E_2)$, for some values of $u$. The values $z(E_1)$ and $z(E_2)$ are 
obtained from the finite-size  extrapolations, where we use in \rf{e4.1} 
the first gap ($E_1$) and the second gap ($E_2$), respectively. The 
differences  
 among $z(E_1)$ and $z(E_2)$ give us an idea of the accuracy of the 
predictions. For $u\geq 5$ we only calculate $z(E_2)$ due to numerical 
instabilities. 
\begin{table*}[htp]
\begin{center}
\begin{tabular}{lcccccr}
\cline{1-7}

   $u$        & 1 & 1.135 & 1.5 & 2 & 6 & 10 \\ \hline 
  $z(E_1)$ & 1.009 & 0.842 & 0.731 & 0.666 & - & - \\
  $z(E_2)$ & 1.009 & 0.900 & 0.756 & 0.642 & 0.184 & 0.030 \\
\hline
\hline
\end{tabular}
\end{center}
\caption{ Estimated vales of the dynamical critical exponent. The 
values $z(E_1)$ and $z(E_2)$ are obtained from the extrapolation  ($L\to \infty$) (see Eq.~\rf{e4.1}) where the first and second gap were taken, respectively.}
\label{table1}
\end{table*}

A possible way to calculate the dynamical critical exponents using  lattice 
sizes $L>30$ is from the time evolution  of some  observable. For example 
taking as observable the average height $h(L,t)$ at time $t$, we do expect 
the general time and size dependence:
\be \label{e4.2}
\frac{h(L,t)}{h(L,(\infty)} -1 = L^{z_2} f(t/L^{z_1}).
\ee

\begin{figure}
\begin{subfigure}{.5\textwidth}
\centering
\includegraphics[angle=0,width=0.8\textwidth] {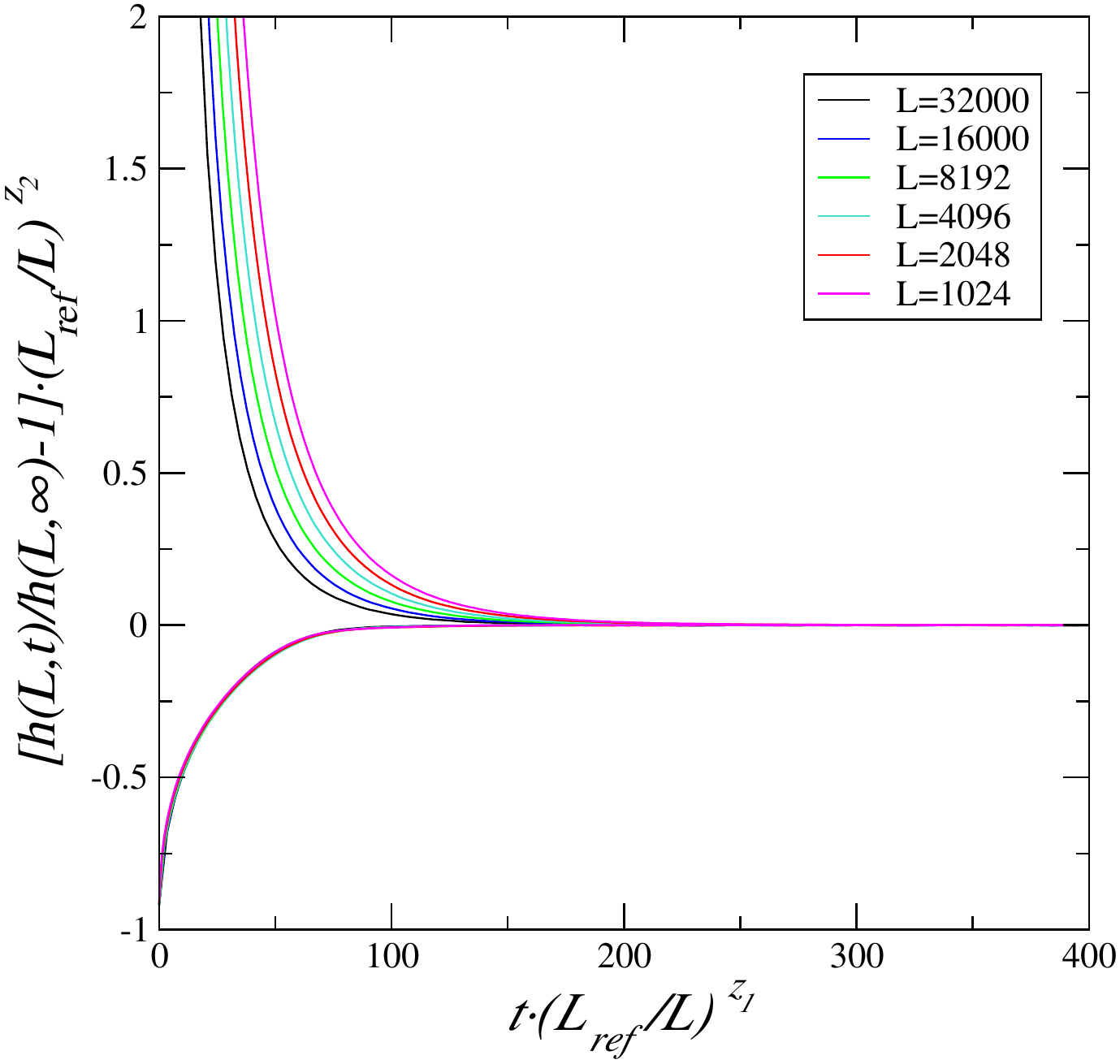}
\label{fig9a}
\end{subfigure}
\begin{subfigure}{.5\textwidth}
\centering
\includegraphics[angle=0,width=0.8\textwidth] {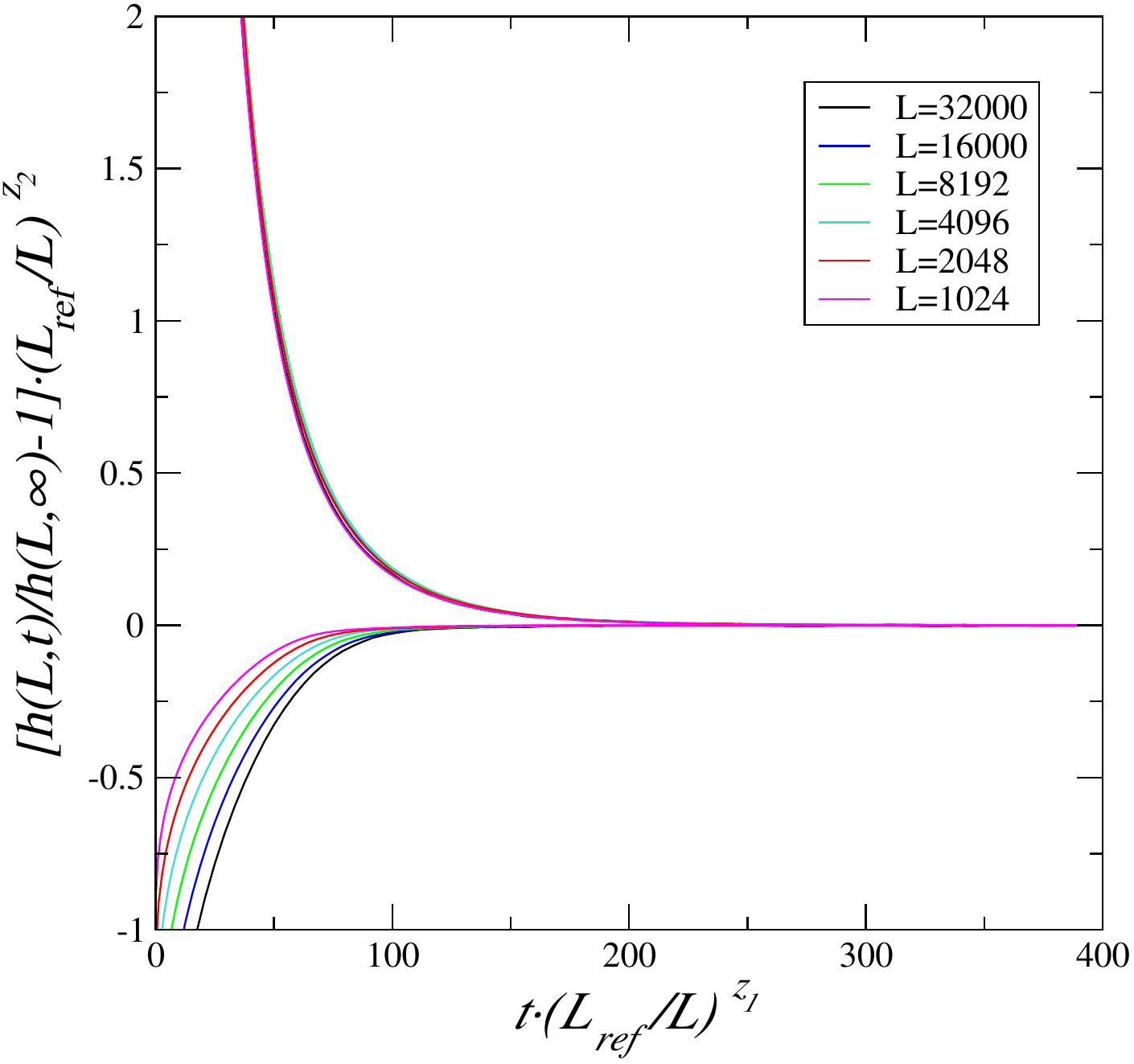}
\label{fig9b}
\end{subfigure}
\caption{
Time evolution curves ($h(L,t)/h(L,\infty) -1$), for the RPM with open boundary 
condition and several lattice sizes. In the top (bottom) curves the initial 
configuration is the pyramid (substrate). A data collapse of the curves, 
following Eq.~\rf{e4.2}, is obtained in (a) and (b) by considering as the 
initial configuration the substrate and the pyramid, respectively. For all 
the curves the harmless parameter $L_{\scriptsize{\mbox{ref}}}=32000$ is used 
to better visualize the figures, and $1.6 10^5$ independent runs were averaged.} 
\label{fig9}
\end{figure}

In general, using different initial conditions we may find distinct pairs of 
the exponents ($z_1,z_2$). If we reach an asymptotic regime where we obtain 
the same exponent $z_1$, for any initial condition, then $z_1$ is the 
dynamical critical exponent. If the system is critical and we do not find 
the same value of $z_2$, for distinct initial conditions, we have the effect 
known as {\it critical initial slip}, as seen in \cite{slip}. In fact for $u>1$ 
this effect is present in the RPM, probably due to its nonlocal dynamical 
processes. In Fig.~\ref{fig9}a and \ref{fig9}b we show the time evolution 
of the average heights \rf{e4.2}, by taking as the initial configuration 
the substrate and the pyramid ones, respectively. We obtain a quite good 
collapse of the curves for the several lattices with 
$(z_1,z_2)=(0.361,0.021)$ and 
$(z_1,z_2)=(0.415,0.282)$  in Figs.~\ref{fig9}a and \ref{fig9}b, 
respectively. The time interval that would give us the same value of $z_1$ 
for both initial conditions happens only at large time  and $h(L,t)/h(L,\infty) -1$ is 
negligible, preventing a reasonable prediction for $z$. Other observables, like 
the number of clusters, also show the same critical initial slip effect. 

Due to this memory effect, in order to get a reasonable estimator for the 
dynamical critical exponent, we should relate this exponent with an 
observable that can be measured directly at the stationary state. In the 
case of periodic boundaries this observable does exist. We conjecture that at 
the stationary regime the average height behaves, apart from a constant, as 
\be \label{e4.3}
h_l(t) = L^{\alpha} f(t/L^z) + v_{\infty} t,
\ee
where $v_{\infty}\equiv v(t\to\infty, L\to \infty)$ is the bulk limit 
velocity where the stationary height grows, and $\alpha$ is the roughness 
critical exponent. The exponent $\alpha$ appears in \rf{e4.3} due to the 
self-affinity of the surface in the stationary state.

\begin{figure}
\centering
\includegraphics[angle=0,width=0.4\textwidth] {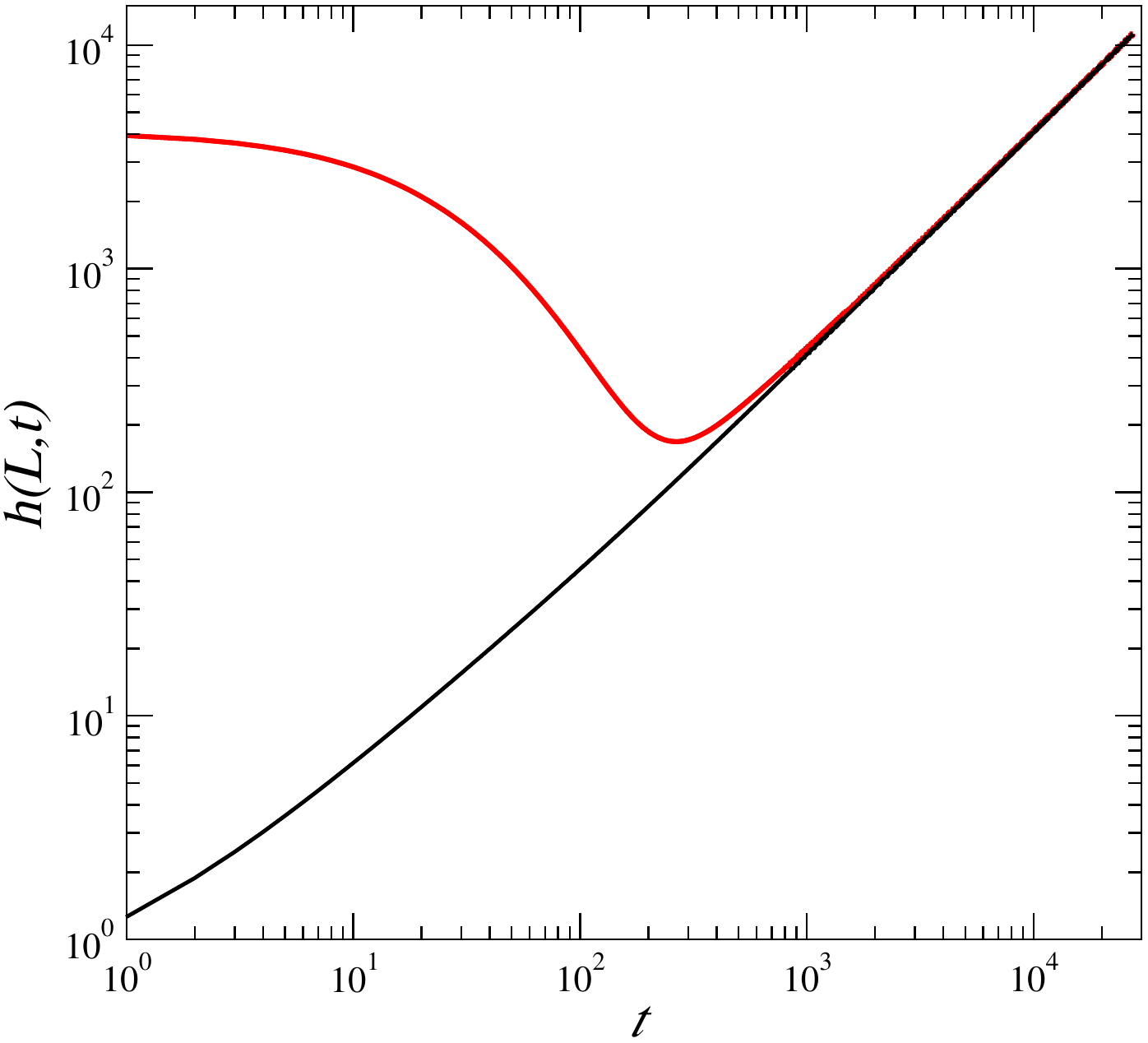}
\caption{
The average height $h(L,t)$ of the RPM with parameter $u=100$ and periodic boundary 
conditions. The initial configuration are the substrate (black) and the 
pyramid  
(red). The lattice size is $L=16,384$ and the number of samples in the Monte 
Carlo simulations 
is 10,048.}
\label{fig10}
\end{figure}

The roughness of a profile is defined as \cite{roughness-definition}
\be \label{e4.4}
\omega(L) = \left[ \frac{1}{L} \sum_{j=1}^{L} (h_j - \bar{h})^2 \right]^{1/2}, 
\ee
where $\bar{h} = \sum_{j=1}^L h_j$. In a self-affine profile, the exponent 
$\alpha$ give us the change in the roughness $\omega(L)$ due to a scale 
dilation ($L \to bL$): $\omega(bL) = b^{\alpha}\omega(L)$. In  the stationary state the 
velocity that the surface grows, for a given lattice size $L$, is given 
from \rf{e4.3} by
\be \label{e4.5}
v(t,L) \equiv \frac{\partial h(L,t)}{\partial t} = \frac{1}{L^{z-\alpha}} g(t/L^z) + 
v_{\infty}.
\ee
As an illustration we show in Fig.~\ref{fig10} the time evolution of the 
average height for the RPM, with $u=100$, $L=16384$ and periodic boundaries. 
The black (red) curve is obtained when the system initiates in the 
substrate (pyramid) configuration. We see from this figure that 
$v(t\to \infty,L)$ coincides for both initial conditions, avoiding thus the 
critical initial slip effect that appeared in other measures. 

It is also interesting to mention that in \cite{corr} it was observed that 
at the roughness transition point ($\alpha=0$) of a particular model \cite{kert},
 the 
growth velocity deviates from its maximal value, at $L\to \infty$, as $L^z$ 
in agreement with \rf{e4.5}. Another example happens in the ASEP where the 
current (related to the velocity in an equivalent growth model) increases as
$J_L-J_{\infty} = a/L$, in agreement with \rf{e4.5} since in this case 
$z=3/2$ and $\alpha =1/2$. 

\begin{figure}
\centering
\includegraphics[angle=0,width=0.4\textwidth] {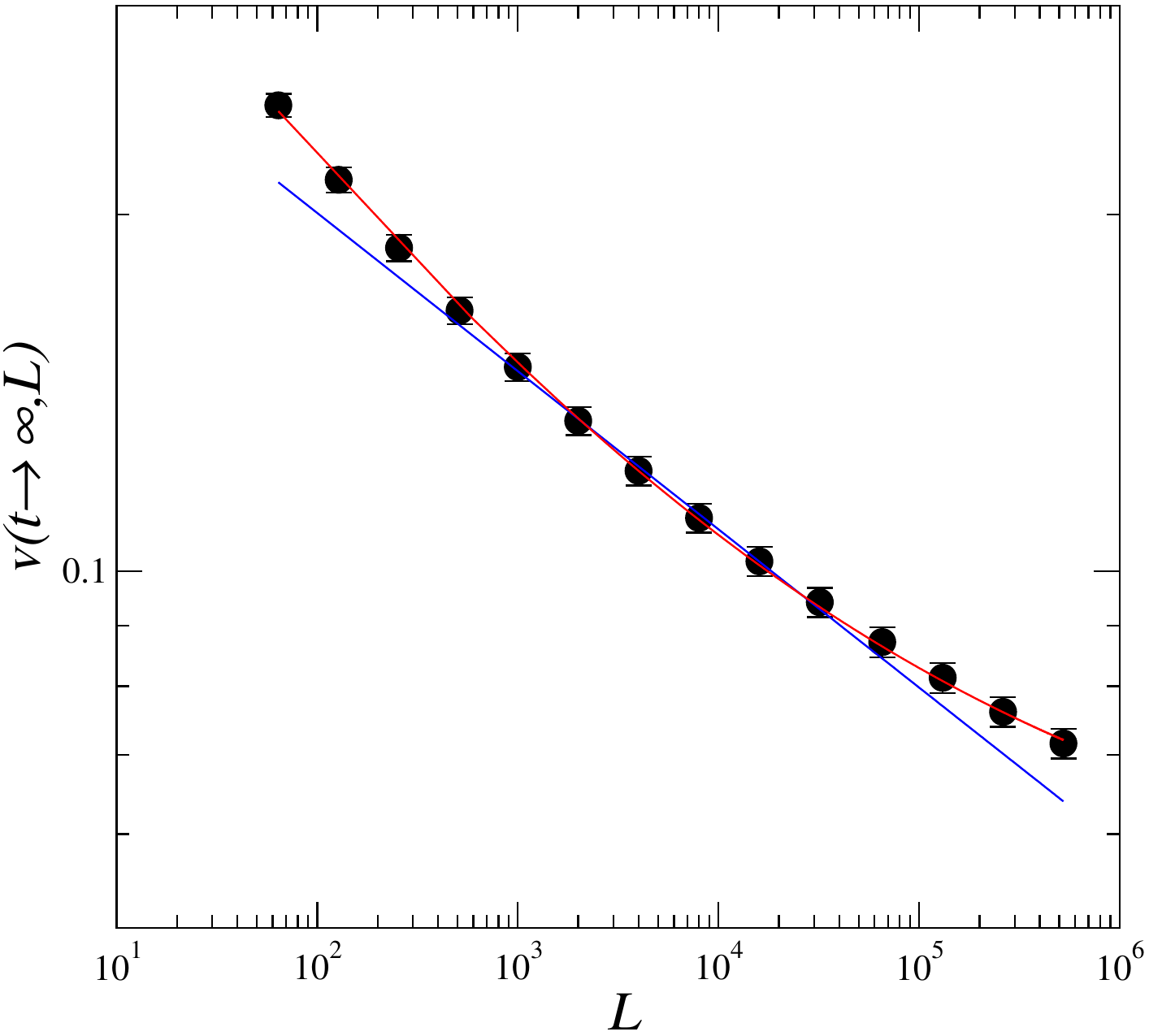}
\caption{
The grow velocity at the stationary regime of the periodic RPM with $u=5$ 
and lattice sizes up to $L\sim 500,000$. The blue curve is the linear 
fit obtained only considering $1,000\leq L \leq 18,000$.}
\label{fig11}
\end{figure}

\begin{figure}
\centering
\includegraphics[angle=0,width=0.4\textwidth] {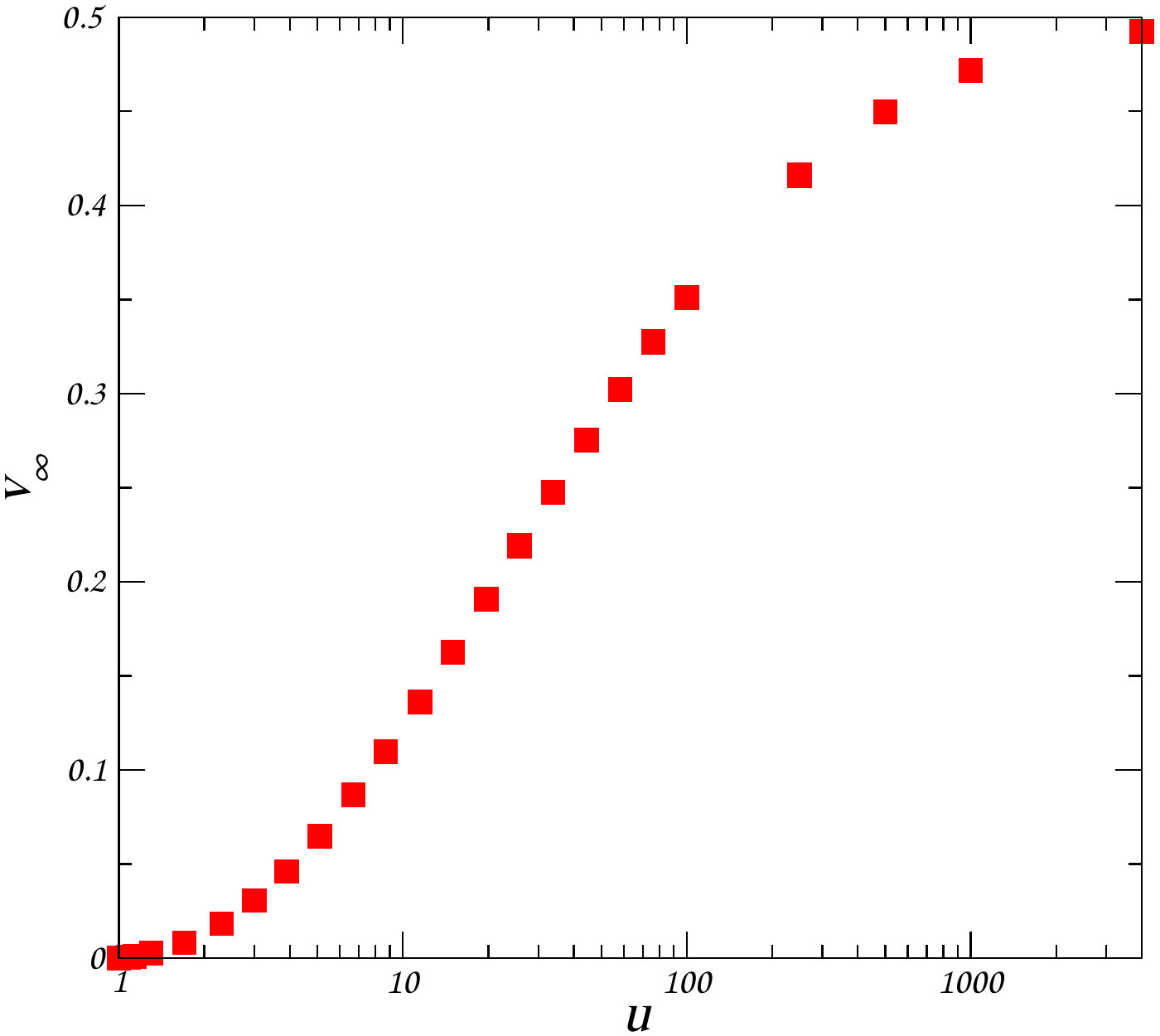}
\caption{
The limiting grow velocity $v_{\infty}= v(t \to \infty,L \to \infty)$ in the 
stationary state of the periodic RPM, as a function of the parameter $u$. 
The results were obtained by averaging 10,048 Monte Carlos samples.}
\label{fig12}
\end{figure}

\begin{figure}
\centering
\includegraphics[angle=0,width=0.4\textwidth] {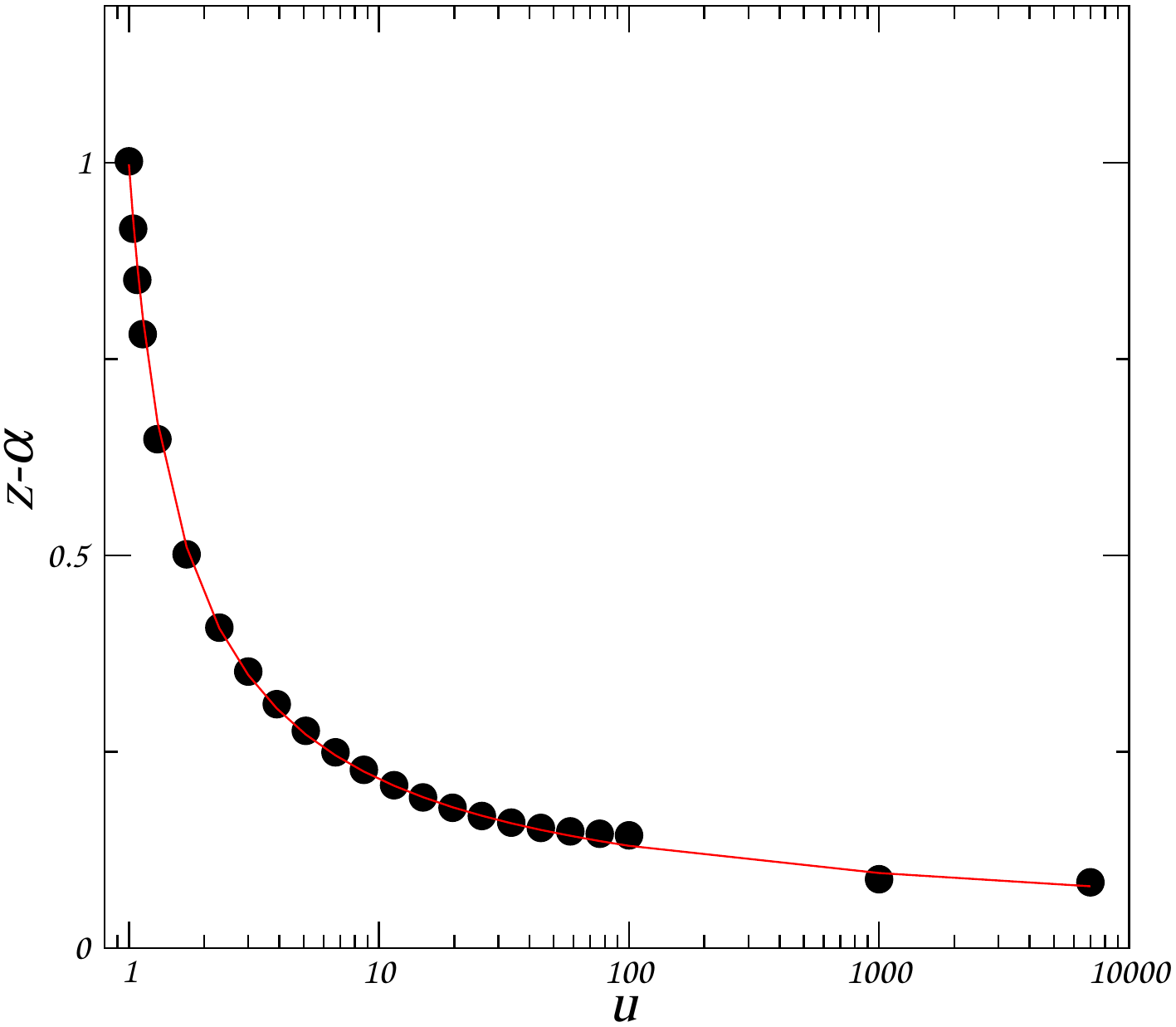}
\caption{
The exponent $z-\alpha$, as a function of $u$, for the RPM with periodic 
boundaries (see Eq.~\rf{e4.5}). The values were obtained by fitting the 
stationary grow velocity $v(t \to \infty,L) = AL^{z-\alpha} + v_{\infty}$. The red 
line is the fitted curve $g(u)= 0.49/(\ln u +0.42)^{0.82}$. The maximum error 
is 9\%, and the number of Monte Carlo samples for each lattice size is $10,048$. The
lattice sizes are $2^6 \leq L \leq 2^{14}$ for all the values of $u$, except $u=1000$ 
and $u=7000$ where $2^6\leq L\leq 2^{19}$.}
\label{fig13}
\end{figure}

Previous studies of the RPM with periodic boundaries \cite{RPM4}, based 
on lattice calculations up to lattice size $L\approx 18000$, indicate that 
the current, in the particle-vacancy representation of the model, or 
equivalently the growth velocity in the  height representation, vanishes for 
$L \to \infty$. However the calculations presented in this paper for larger 
lattices indicate that in fact for $u>1$ the velocity in the bulk limit is 
nonzero. As an example we show in Fig.~\ref{fig11} the growth velocity for 
the model with parameter $u=5$ and lattice sizes up to $L \sim 500,000$. We 
also show in the figure the fitted curve (blue) for the lattice 
sizes  $1000<L<18000$, used in \cite{RPM4} 
that indicates the vanishing of the current (velocity) as $L \to \infty$. 
The curvature shown in the figure indicates that the velocity will saturate 
in a nonzero value as $L\to \infty$. If we adjust the stationary velocity 
of Fig.~\ref{fig11} as $v(t \to \infty,L) = A/L^{z-\alpha} +v_{\infty}$ we 
obtain for $u=5$, $z-\alpha=0.242$ and $v_{\infty} =0.0255$. Repeating the 
procedure for other values of $u$ we obtain the limiting velocities 
$v_{\infty}=v(t \to \infty, L \to \infty)$ shown 
in Fig.~\ref{fig12}, and the exponents $z-\alpha$ shown in Fig.~\ref{fig13}. 
As we can see $z-\alpha$ decreases drastically as $u \to \infty$. A fitted 
curve for the values $z-\alpha=0.49/(\ln u +0.42)^{0.12}$ is shown in red 
in Fig.~\ref{fig13}. The limiting behavior when $u\to \infty$ is distinct 
from the one at $w=1/u=0$. For $1/u=0$ the model is equivalent to the TASEP, 
and the boundary condition is important. For the open case, since the 
particles do not travel after reaching the lattice border, the model is 
noncritical, while in the periodic case the model is critical and belongs to 
the KPZ universality class where $z=3/2$ and $\alpha=1/2$, therefore 
$z-\alpha=1$. In the next section we are going to see how the profiles 
expected on the KPZ dynamics appear in the short length scale of the RPM with 
large values of $u$. 

\begin{figure}
\centering
\includegraphics[angle=0,width=0.4\textwidth] {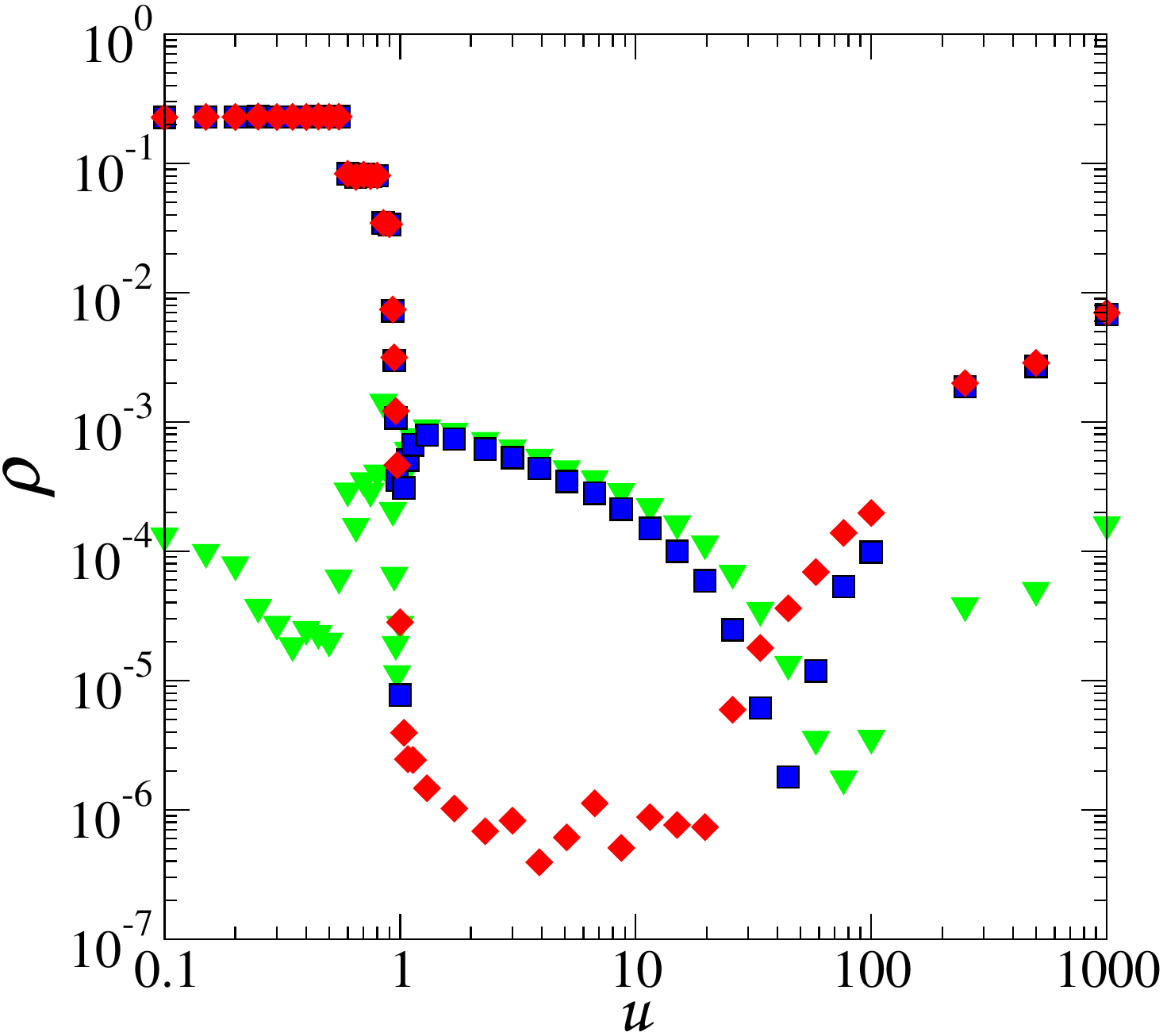}
\caption{
The estimator 
 $\rho =1-(\mbox{correl}(\kappa_1,\hat{\kappa}_1))^2$, as a function of the 
parameter $u$, for the  fittings of the square of the rugosity 
$<\omega^2>$ (see Eq.~\ref{e4.4}) of the RPM with periodic boundaries in the 
stationary regime. The best fittings are the ones with smaller values of 
$\rho$. The point in red, green, and blue are the ones where the adjusted 
behavior, for the windows of size $\ell$ are 
$f_{b>0} = a\ell^b + c$ (with $b>0$),
$f_{b<0} = a\ell^b + c$ (with $b<0$) and 
$f_{\scriptsize{\ln}}=\ln \ell +c$, respectively (see also the text).}
\label{fig14}
\end{figure}

\begin{figure}
\centering
\includegraphics[angle=0,width=0.4\textwidth] {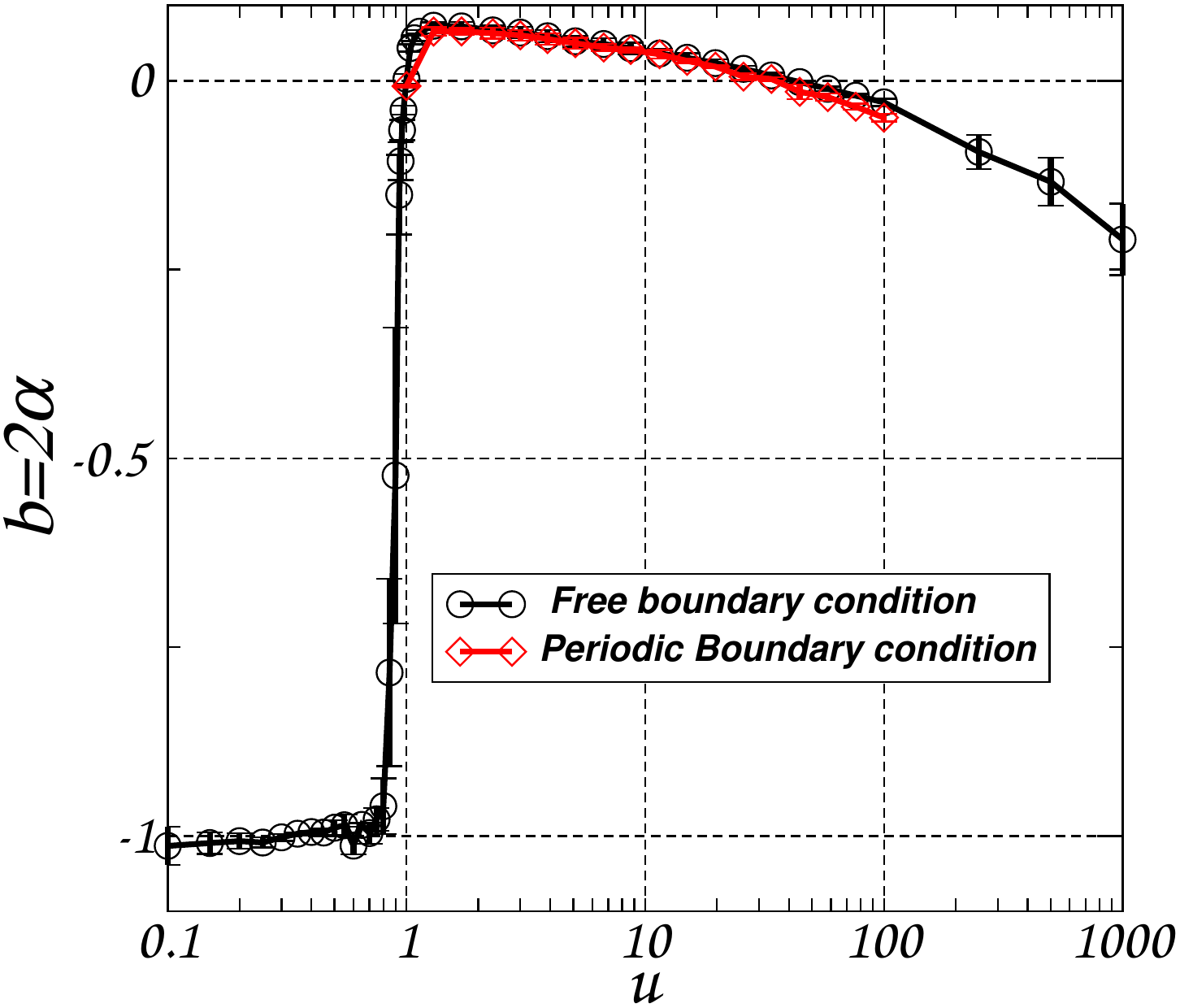}
\caption{
The exponent $b=2\alpha$ as a function
of $u$. For positive values $\alpha$ is the roughness critical exponent. 
They  are obtained by choosing, in the height 
profiles, windows of size $\ell$ and fitting their average to the 
function $<\omega^2(l,L)> =a \ell^b +c$. The curve in black (red) 
is for periodic (open) boundary conditions, and $\ell=L$ ($1 <<\ell<<L$).}
\label{fig15}
\end{figure}

In order to finish our calculation of the critical exponent $z$ we need to 
calculate the roughness exponent $\alpha$. This exponent was obtained by 
considering windows of the profiles $\{h_i\}$  of size 
$\ell$, localized at the center of the lattice (size $L$). We fit the first 
cumulant $\kappa_1(\ell,L) = <\omega^2(\ell,L)>$ of the roughness \rf{e4.4} 
to the three distinct behaviors: $f_{b>0} (\ell) = a \ell^b +c$, 
for $b>0$, $f_{b<0}(\ell) = a\ell^b +c$, for $b<0$, or 
$f_{ln}(\ell) = a\ln \ell +c$. Among these fits the best one is the one 
where $\rho =1-(\mbox{correl}(\kappa_1,\hat{\kappa}_1))^2$ has the smallest 
value. The $\mbox{correl}(\kappa_1,\hat{\kappa_1})$ is the correlation among the 
original data of $\kappa_1$ and the fitted ones. In Fig.~\ref{fig14} we show 
the values of $\rho$ for the model with periodic and free boundary conditions. 
The results were obtained for lattice sizes $2^{10} \leq L \leq 2^{14}$ and 
for the values of $1 \leq u \leq 1000$. We see from these curves that 
$b<0$ for 
$u<1$ and $u \gtrsim 40$ (green triangles) and  $b>0$ for $1<u \lesssim 40$ 
(red squares). We also observe that  the logarithmic fittings (blue squares) are 
reasonable only at $u=1$ and $u\sim 40$. Using these fits we obtain the values 
of $2\alpha$ shown in Fig.~\ref{fig15} (black curve). 
We also include in Fig.~\ref{fig15} the values of $2\alpha$ (red curve) obtained 
for $0.1 <u <100$ when the RPM is defined in a lattice with free ends. 
 We see an agreement of the estimated 
values by imposing the two distinct  boundary conditions. 

These results indicate that for $u>1$ the RPM has two distinct phases. The 
transition at $u=u_0=1$ can be seen as a roughening transition. For $u<u_c 
\approx 40$ we have a rough phase with $\alpha>0$, while for 
$u>u_c \approx 40$ the surface is not rough ($\alpha <0$). The values of 
$\alpha$ in the rough phase are small and vary continuously with $u$. Since 
for $1<u<u_c$ the obtained values of $\alpha$ are small, we should convince 
ourselves that this variation in not just a finite-size effect, and 
$\alpha$ could be constant in the whole phase. However performing the 
evaluation of $\alpha$ by selecting windows of the profiles on distinct ways 
we obtain almost the same results presented in Fig.~\ref{fig15}, indicating 
that in this rough phase $\alpha$ varies continuously. In order to better 
understand the nature of the two critical phases we are going to calculate, 
in the next section, the structure function of the height profiles in both 
phases. 

\section{The structure functions of the height profiles for \\$u>1$} \label{section5}

  The results of the last section indicate that the RPM has, for $u>1$, two 
distinct phases. For $1<u<u_c\approx 40$ the surface profiles are rough with 
roughness exponent $\alpha >0$, while for $u>u_c$ the exponent $\alpha<0$. In 
this section we calculate the structure functions of the height profiles.

The structure function \rf{e2.6} in the stationary regime has the leading 
behavior, for small values of $k$ or large values of $\lambda = 2\pi/k$:  
\cite{Lie}
\be \label{e5.1}
S(k,t\to \infty) = S(k) \sim k^{-(1+2\alpha)},
\ee
that depends on the roughness exponent $\alpha$.

Let us calculate initially the expected behavior of $S(k)$ at the limiting 
cases $1/u =0$ and $u=1$. At $1/u=0$ the RPM with periodic boundaries recovers 
the critical TASEP, or equivalently the single step growth 
model \cite{TASEP1,Huse}. The height-height correlation function in the 
stationary state ($t \to \infty$) of this last model, for heights at position $i$ and $j$, has 
the general behavior
\be \label{e5.2}
<(h_j -h_i)^2> - [<h_j>-<h_i>]^2 = \frac{D}{2\nu}(j-i)^{2\alpha}, \quad 
2\alpha=1,
\ee
where $h_i=h(i,t\to \infty)$ and 
 $1<<j-i<<L$. In the mapping with the RPM we should take
 $D=2$ and $\nu=1$. Then from \rf{e5.1} we obtain, at $1/u=0$, 
\be \label{e5.3}
S(k) \sim \frac{1}{k^2},
\ee
for large values of $\lambda=2\pi/k$.

\begin{figure}
\centering
\includegraphics[angle=0,width=0.4\textwidth] {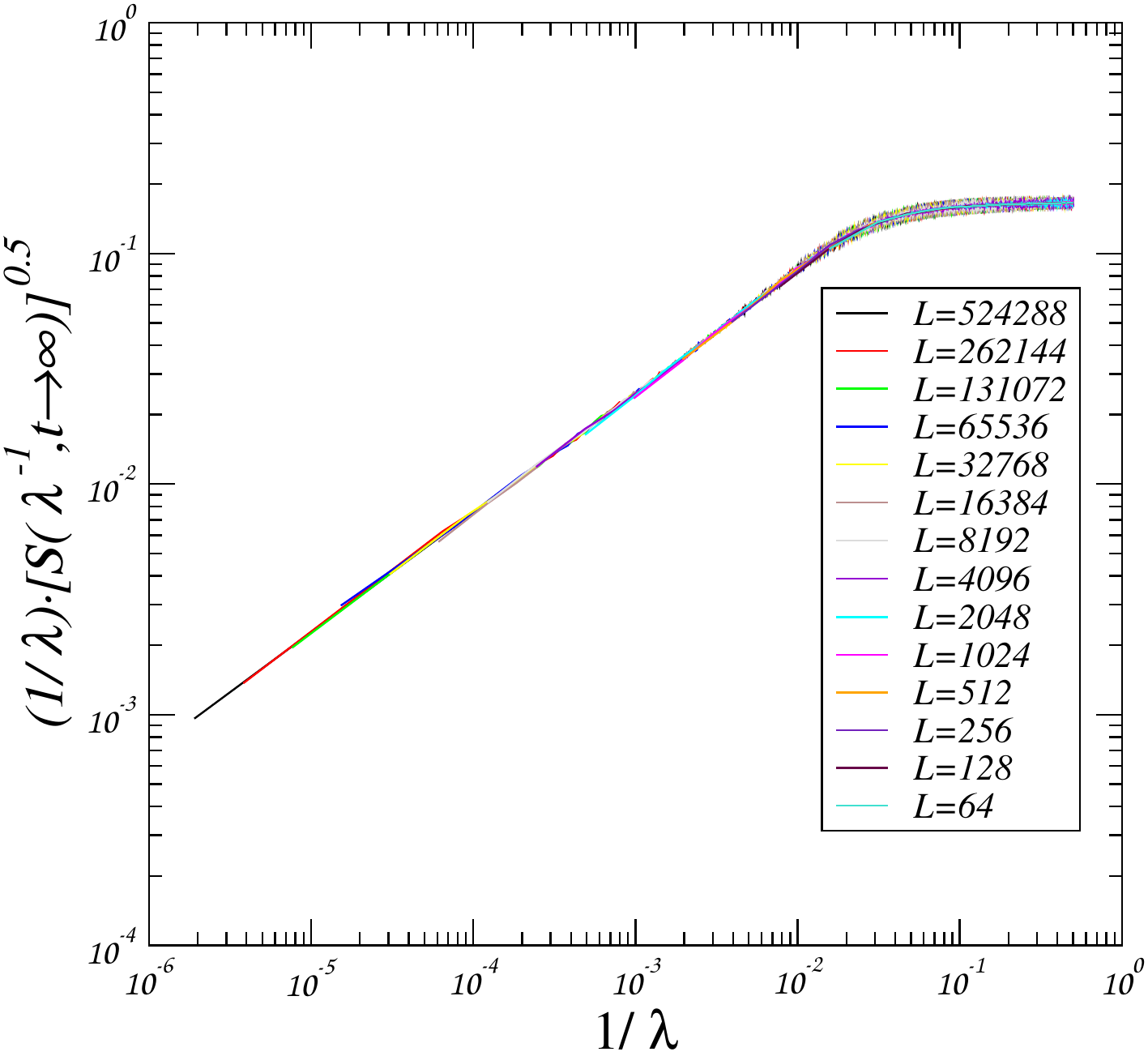}
\caption{
The structure function $S(1/\lambda,t \to \infty)$, as a function of $1/\lambda$ for 
the height profiles of the RPM on its stationary regime. The parameter $u=100$ and 
the several lattices are shown in the figure. It was considered 14 samples in 
the Monte Carlos simulation, for each lattice size.}
\label{fig16}
\end{figure}

In Fig.~\ref{fig16} we show the stationary structure function $S(1/\lambda)$ 
 of the RPM at stationary time for $u=100$ and several lattice sizes. We 
clearly see a crossover region at short length scales 
$\lambda = 2\pi/k \lesssim 70$. At these short scales $k^2S(k)$ is constant 
and the profiles are similar as the ones of the single step height model, 
or the TASEP, where the roughness exponent is $\alpha=1/2$. It is interesting 
to notice that the crossover regions is almost lattice size insensitive. 

The other interesting limit of the RPM is the conformal invariant point $u=1$. 
At this point by exploring the underlying conformal symmetry of the model it 
was obtained in \cite{corr}, that at the stationary time
\be \label{e5.4}
<h^2> -<h>^2 = \frac{1}{L}\sum_k S(k) \sim \frac{2\sqrt{3} \pi -9}{\pi^2} \ln L,
\ee
where $h=h(L,t\to \infty)$ is the height of a configuration given in \rf{e2.4}.

Using \rf{e5.1} and \rf{e5.4}  we obtain that $\alpha=0$ and 
\be \label{e5.5}
S(k)\sim \frac{2\sqrt{3}\pi-9}{\pi}\left( \frac{1}{k}\right).
\ee
In  section 3 it was shown in Fig.~\ref{fig7}, that for $u<1$ there exists 
a crossover region $\lambda <  \lambda_c$, that although the system is 
noncritical it behaves as in \rf{e5.5}.

\begin{figure}
\centering
\includegraphics[angle=0,width=0.4\textwidth] {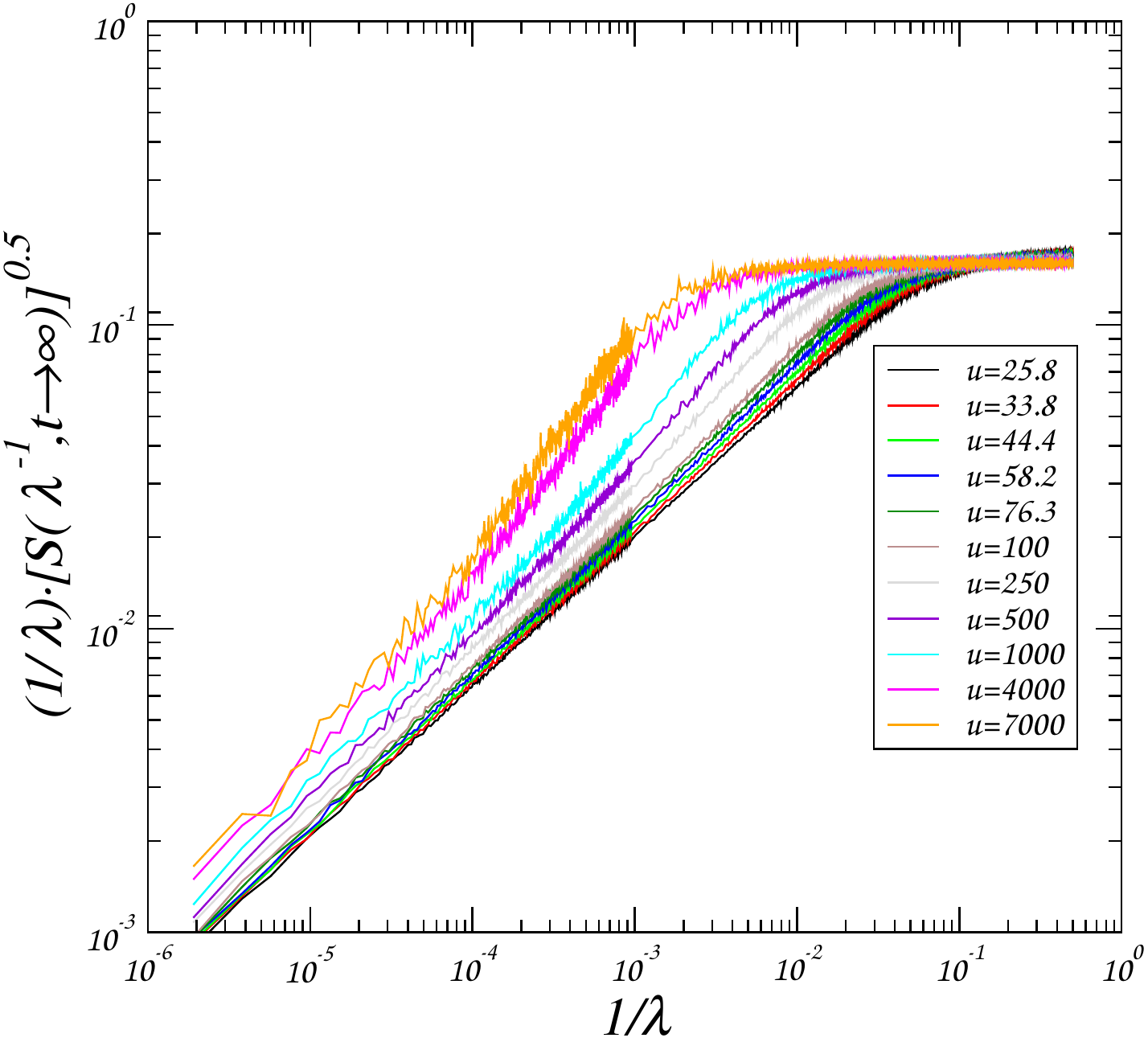}
\caption{
The structure function $S(1/\lambda,t \to \infty)$ of the height profiles of 
the RPM with several values of parameter $u$. The lattice size is $L=524,288$ and 
periodic boundary conditions were used.}
\label{fig17}
\end{figure}

\begin{figure}
\centering
\includegraphics[angle=0,width=0.4\textwidth] {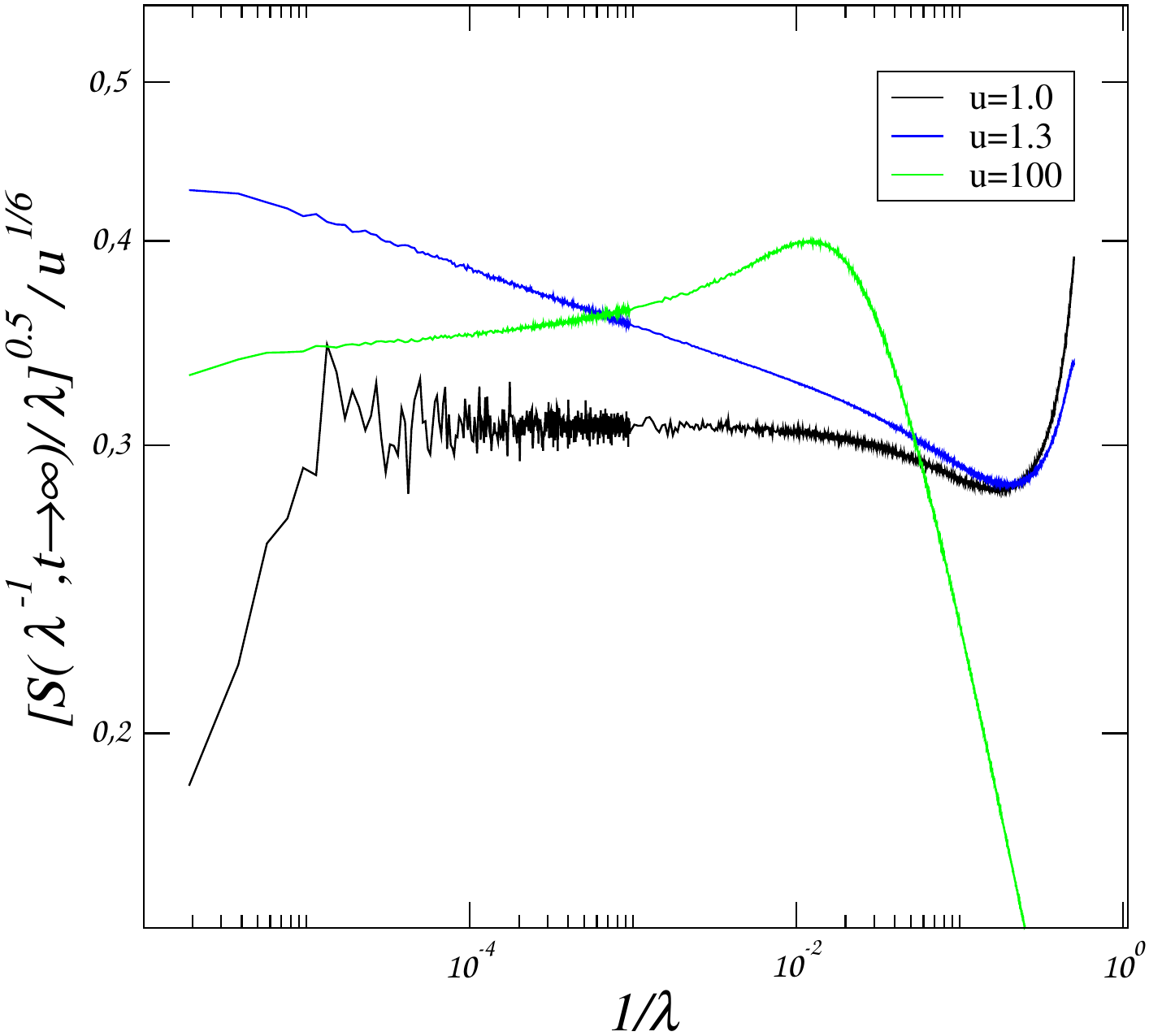}
\caption{
The structure function $S(1/\lambda,t \to \infty)$ of the height profiles of 
the RPM with  parameters  $u=1,1.3$ and $100$.  The lattice size is $L=524,288$ and 
periodic boundary conditions were used. Notice that a multiplicative factor 
$u^{1/6}$ was inserted  in the vertical axis in order to better visualize the 
three curves.}
\label{fig18}
\end{figure}

In Fig.~\ref{fig17} we show $S(k)$ for a fixed large lattice size 
$L=524,288$ and several values of $u>1$. We clearly see in this figure that 
even for $u \approx 25$ there exists the crossover region 
$\lambda <\lambda_c$ where the model exhibits the KPZ behavior ($\alpha =1/2$).
This crossover length $\lambda_c(u)$ increases with $u$, being infinite at 
$1/u=0$. The figure show us that no matter how large is the system the 
crossover length $\lambda_c(u)$ is finite, and for sufficiently large scales 
($\lambda >>\lambda_c(u)$), the model has a distinct $u$-dependent behavior. 
This crossover region, with the KPZ behavior, explains the difficulty in 
measuring the roughness exponent $\alpha$ directly in the real space, for 
large values of $u$. As we increase $u$ we need to consider also larger 
lattices in order to obtain reasonable estimates for $\alpha$. However, in the 
reciprocal space we can separate the behavior at large scales from the 
small ones. Considering only the large scales we  obtain a 
good estimate for $\alpha$. To 
illustrate the sensibility of the structure function, 
even for 
small values of $\alpha$, we show them in Fig.~\ref{fig18}, for the lattice size 
$L=524,288$ and for $u=1$, $1.3$ and $u=100$.  We clearly see that the large 
scales gives $\alpha=0$, $\alpha>0$ and $\alpha<0$, respectively.

\begin{figure}
\centering
\includegraphics[angle=0,width=0.4\textwidth] {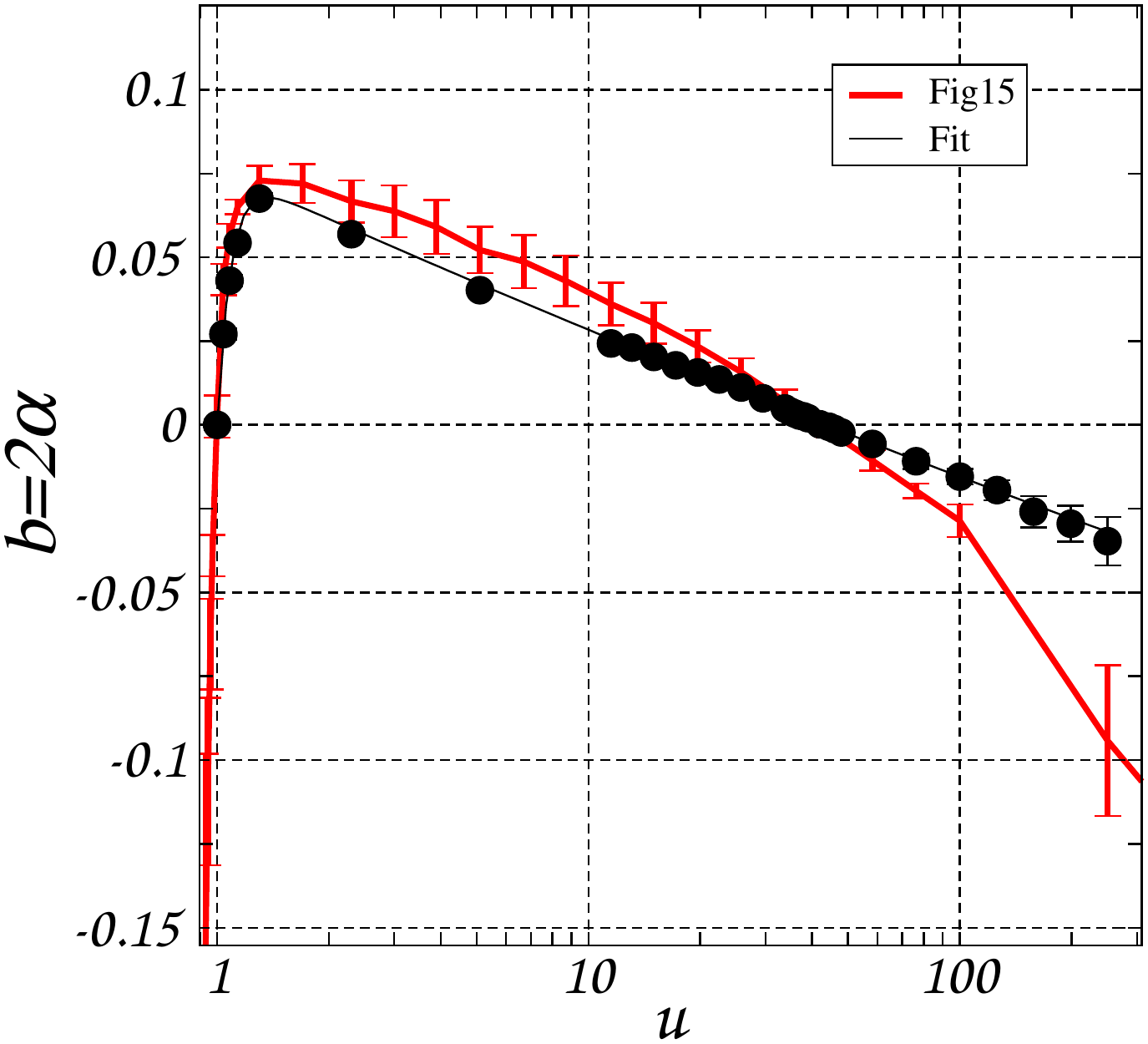}
\caption{
The estimated values of the exponent $b=2\alpha$, as a function of $u$. The black 
dots are the results obtained from the structure function behavior at large 
scales and the red curve are the ones shown in Fig.~\ref{fig15},  obtained 
 by adjusting the behavior of the square of the rugosity  in windows  of 
several 
sizes. The continuum black curve is a fitting one 
(see the text).}
\label{fig19}
\end{figure}

   For small values of $k$, from \rf{e5.1},  we estimate the exponent $\alpha$ 
for several values of $u\geq 1$ ($1/u=w<1$). The estimates obtained for the 
lattice size $L=2^{19}=524,288$ are the black dots in Fig.~\ref{fig19} (red curve). For the
sake of comparison we also show in this figure the estimated values of 
$\alpha$ obtained in Fig.~\ref{fig15}. The results obtained from the structure 
function calculations, that we believe are more precise, as compared with the 
one obtained in last section, give us the fitted curve (continuous black curve 
in Fig.~\ref{fig15}):
\be \label{e5.6}
2\alpha= \frac{ b^{\frac{1}{u^c}}}{b+a\ln u} -1, 
\ee
with $a=0.0188(1)$, $b=0.9290(3)$ and $c=12.2(4)$. These results confirm 
the ones obtained previously in the last section and indicate that the system 
is in a rough phase ($\alpha \geq 0$) for $1\leq u \leq u_c \simeq 40$ and 
for $u>u_c$ the system is in phase that at large scales is flat ($\alpha >0$).

\section{Summary and conclusions} \label{section6}

In this paper we present an extensive study of the raise and peel model (RPM) 
with open and periodic boundary conditions. The dynamical processes defining 
the time evolution of the RPM are local adsorptions and nonlocal desorptions 
 (avalanches). The relevant parameter of the model is the ratio $u$ 
($0\leq u$ ) among the adsorption  and desorption rates. At $u=1$ the 
RPM give us the first example of a conformally invariant stochastic model. 
Previous studies of the RPM \cite{RPM1,RPM2} indicate that for $u<u_0$ the 
model is massive (noncritical) while for $u>u_0$ the model is in a critical 
phase. The critical point $u_0$ is not simple to evaluate, although previous 
results indicate that $u_0=1$. In this paper we obtain a precise evaluation of
$u_0=1.000(1)$. This was done by calculating the structure function of the 
height profiles in the reciprocal space. Our results explains the difficulty 
in evaluating $u_0$ by direct measures of observables. There exists for 
$1 > u\gtrsim 0.4$, where the model is clearly massive, a crossover region 
at short scales $\lambda<\lambda_c(u)$, where the model has an effective 
behavior similar as the one at the critical conformal invariant point $u=1$. 
This crossover length $\lambda_c(u)$ grows and diverges at $u=u_0=1$. If we 
do not consider only the large scales $\lambda >\lambda_c(u)$, as we do in 
direct measurements of observables, when $u$ is close to $u_0=1$ the 
crossover behavior dominates, preventing us to obtain the large-distance 
physics of the model. 

Since at $u_0=1$ the roughness exponent $\alpha=0$ is known 
exactly \cite{corr}, we establish that the RPM has a roughness transition at 
the conformal invariant point $u=u_0=1$.

For $u>1$ previous results \cite{RPM2,RPM4} indicate that the RPM is in a 
critical phase, with self-organized criticality having a dynamical critical 
exponent $z(u)$ that decreases continuously with the parameter $u$. On the 
other hand, as noticed in \cite{Jara}, in the limit $1/u=0$ the RPM  is 
exactly mapped into the TASEP, a model with quite distinct properties, 
if the boundary conditions are taken to be open (free) or closed (periodic). 
For open boundary conditions the model is massive and for the periodic ones 
the model is critical and belongs to the KPZ critical universality with 
$z=3/2$ and $\alpha=1/2$. The estimated value $z(u\to \infty) =0$ 
of the dynamical critical exponent of the RPM was obtained by applying 
open boundary conditions \cite{RPM2}. The above mapping of the RPM with the 
TASEP clearly rises the question if the physical behavior of the RPM, for 
large values of $u$, is also boundary condition dependent as happens at 
$1/u=0$. Our results indicate that this is not the case. Although $z(u) \to 0$ 
as $u\to \infty$, with a roughness exponent $\alpha <0$, at the limit $1/u=0$ 
the model has the KPZ exponents $z=3/2$ and $\alpha=1/2$. This is an effect 
of the non locality of the model. As long $1/u\neq 0$ the nonlocal processes 
of the RPM produce the same long-distance behavior in both boundaries. 
Our calculations of the dynamical critical exponents $z(u)$ for the RPM with 
$u>1$ were done initially by exploring the $L$-dependence of the mass gaps of 
the Hamiltonian  on small lattice sizes $L \leq 30$. Better 
estimates of $z(u)$ could be obtained, in principle, from the time evolution 
of observables. Our results show, however, that for quite large times the 
time dependence is distinct for different initial conditions. This effect 
is known as critical initial slip \cite{slip} and forbid us to get reliable 
results since $z(u)$ is small for large values of $u$. In order to avoid this 
large memory effect we need to relate $z(u)$ with a quantity that can be 
measured directly in the stationary state ($t\to \infty$). We found this 
quantity in the periodic case. For this boundary condition the average 
height grows with time and the we relate the growth's velocity of the 
surface with the difference of the critical exponents $z(u)-\alpha(u)$.

We evaluate the roughness exponent $\alpha(u)$ directly from the roughness 
of the surface and surprisingly we found that the region $u>1$ is not a 
single phase but is composed by two critical phases. For $u<u_c \approx 40$ 
the model is rough with a positive exponent $\alpha(u)>0$, while for 
$u>u_c$ the model is not rough having $\alpha(u) <0$. 

In order to better understand the critical phases for $u>1$ we evaluate the 
structure functions of the height profiles in the reciprocal space. These 
calculations show us that as long $u\geq 1$ the KPZ behavior is present in 
the RPM at short length scales. The crossover region, where the model has 
a KPZ behavior, grows with the parameter $u$, becoming infinite only at 
$1/u=0$, where indeed the model is equivalent to the TASEP. Although the 
estimated values of the roughness exponents $\alpha \lesssim 0.02$ 
are small 
the behavior of the structure function at large scales ($k$ small) 
clearly indicates that the phase $1\leq u \leq u_x\approx 40$ is rough 
while the phase $u>u_c$ is flat for the large length scales. 

\section{Acknowledgments}
It is a great pleasure to thank Vladimir Rittenberg for numerous and 
stimulating  discussions throughout the course of this work, and also for 
a careful reading of the manuscript.
  This work was support in part by the Brazilian funding agencies: 
  FAPESP, CNPq and CAPES.


\begin{thebibliography}{99}
	\expandafter\ifx\csname url\endcsname\relax
	\def\url#1{{\tt #1}}\fi
	\expandafter\ifx\csname urlprefix\endcsname\relax\def\urlprefix{URL }\fi
	\providecommand{\eprint}[2][]{\url{#2}}
%
	\bibitem{RPM1}
	De~Gier J, Nienhuis B, Pearce P~A and Rittenberg V 2004 {\em J. Stat. 
 Mech.\/} {\bf 114} 1--35 
%
\bibitem{RPM2}
	Alcaraz F~C, Levine E and Rittenberg V 2006 {\em J. Stat. Mech.} 
{\bf P08003}
%
\bibitem{RPM3}
	Alcaraz F~C and Rittenberg V 2007 {\em J. Stat. Mech.} 
{\bf P07009}\\
	Alcaraz F~C, Pyatov P and Rittenberg V 2008 {\em J. Stat. Mech.} 
{\bf P01006}
%
\bibitem{RPM4}
	Alcaraz F~C and Rittenberg V 2013 {\em J. Stat. Mech.} 
 {\bf P09010} 
%
%
\bibitem{ASEP}
 Derrida B, Domany E and  Mukamel D, 1992 {\it J. Stat. Phys.}
{\bf 69} 667 \\
Derrida B, Evans M R, Hakim V and Pasquier V, 1993 {\it  J. Phys. A}
 {bf 26} 1493
%
\bibitem{Jara} 
Jara D~A~C and Alcaraz~F~C 2017 {\em J. Stat. Mech.} 
{\bf 043205}
%
\bibitem{TASEP1} 
Derrida B, Evans M and Mukamel D 1993 {\em J. Phys. A: Math. Gen.\/} {\bf 26} 4911
%
\bibitem{TASEP2}
Derrida B and Mallick K 1997 {\em J. Phys. A: Math. Gen.\/} {\bf 30} 1031
%
\bibitem{KPZ}
Kardar M, Parisi G and Zhang Y-C 1986 {\em Phys. Rev. Lett.} {\bf 56} 889 \\
Halpin-Healy T and Zhang Y-C, 1995 {\em Phys. Rep.} {\bf 254} 215
%
\bibitem{slip}
Henkel M, Hinrichsen H, L{\"u}beck S and Pleimling M 2008 {\em Non-equilibrium
phase transitions\/} vol~1 (Springer) \\
{\'O}dor G 2008 {\em Universality in nonequilibrium lattice systems\/} (World
Scientific) \\
Grassberger P and Torre A D L 1979 {\em Ann. Phys. } {\bf 122} 373 \\
Jansen H, Schaub B and Schmittmann B 1989 {\em Zeischrift f\"ur Physik B} 
{\bf 73} 539
%
\bibitem{roughness-definition} 
Barab{\'a}si A L and Stanley H E 1995 {\em Fractal concepts in surface
		growth\/} (Cambridge university press)\\
 Reis F A 2001 {\em Phys. Rev. E} {\bf 63} p056116
%
\bibitem{Razumov} 
Razumov A~V and Stroganov Yu~G 2001 {\em J. Phys. A: Math. Gen.} {\bf 34} 3185
%
\bibitem{kert}
Kert\'esz J and Wolf D~E 1989 {\it Phys. Rev. Lett.} {\bf 62} 2571
%
\bibitem{corr}
Alcaraz F~C and Rittenberg V 2015 {\em J. Stat. Mech.} 
 {\bf P11012} 
%
\bibitem{Lie} 
Lie D and Plischke M 1988) {\em Phys. Rev. B} {\bf 38} 4781
%
\bibitem{Huse} 
Huse D A, Henley C L and Fischer D S 1985 {\em Phys. Rev. Lett.} {\bf 55} 2924
%
\end{thebibliography}
\end{document}